\documentclass{article}
\usepackage{graphicx}
\usepackage{amsfonts}
\usepackage{amsfonts,amssymb,amsmath}
\newcommand{\R}{\mathbb{R}}

\begin{document}

\title{Visualizing and Quantifying Impact and Effect in Twitter Narrative using Geometric Data Analysis}

\author{Fionn Murtagh (1), Monica Pianosi (2), Richard Bull (2) \\
(1) School of Computer Science \& Informatics \\
(2) Institute of Energy \& Sustainable Development \\
De Montfort University, Leicester LE1 9BH, UK \\
{\tt fmurtagh@acm.org, rbull@dmu.ac.uk, mpianosi@dmu.ac.uk}}

\maketitle

\begin{abstract}
We use geometric multivariate data analysis which has been termed a methodology
for both the visualization and verbalization of data.  The general objectives are 
data mining and knowledge discovery.  In the first case study, we use the narrative 
surrounding very highly profiled tweets, and thus a Twitter event of significance 
and importance.  In the second case study, we use eight carefully planned Twitter 
campaigns relating to environmental issues.  The aim of these campaigns was to 
increase environmental awareness and behaviour.  Unlike current marketing, political
and other communication campaigns using Twitter, we develop an innovative approach 
to measuring bevavioural change.  We show also how we can assess statistical 
significance of social media behaviour.  
\end{abstract}


\section{Introduction}

The general aim of our work is the ``visualization and verbalization of data''
\cite{greenacre}.  Furthermore, the data here is the narrative of the flow of tweets
(microblogs) in the online (Web 3.0) social medium, Twitter.  

\subsection{Some Current Approaches to Analysis of Twitter Conversations}

The general approach to analysis of Twitter conversations taken in \cite{pearce} 
is based on hashtags (terms preceded with the character ``\#'' which 
can be cross-referenced in Twitter) and users (preceded with the ``@'' character).  
More than 10 connections were required between users (a connection of a user to another 
being the explicit use of each other's ``@'' names) and a graph of such exchanges 
was used for community analysis.  The latter, in the case of \cite{pearce}, was 
primarily aimed at the pro and contra viewpoints relative to climate change.
Based on such polarization of views, and greater prevalence of tweeting in the 
unsupportive-to-supportive direction (relative to action to counteract 
climate change), it was nonetheless concluded that more work was required:
 ``Content analysis of the tweets
could be a possible qualitative approach that could shed light on [...] and provide
new knowledge about the content of conversational connections discovered ...''.  In 
our work in this article, we look at the conversational connections starting from 
what is aiming at being an initiating, instigational and influencing tweet.  

In \cite{chew}, Twitter-based behaviour (relating to the 2009 H1N1 swine flu) 
was subjected to 
content analysis that included analysis of retweets, seeking particular words and 
phrases, and manual labelling for content and sentiment characterization, followed
by analysis of that. Such work was carried out by querying the Twitter data.  The 
queries were sophisticated with many boolean connectives (``and'', ``or'', etc.).  
Rather than a querying, matching and supervised approach such as this, and used 
in general for sentiment 
analysis, our work in this article will be data-driven and unsupervised.  We will 
map out the underlying semantics of our social media data through the text.  The 
text used provides the ``sensory surface'' \cite{mckee} of the underlying semantics.  

Social media monitoring was originally adopted by public relations and advertising 
agencies, who used it as a means to identify negative comments posted on the web 
about their clients \cite{barker}.  It is defined as the activity of observing and 
tracking content on the social web. Each activity on social media has an outcome, or 
effect, which can be measured by observing and then quantifying specific behaviours.  
Effects can be one of the following: retweets, mentions, favourites, follows, likes, 
shares, comments, sentiment.

Social media are used by companies and public relations agencies, by local and 
central governments, who all seek to evaluate the use of social media channels as a 
communication or engagement tool.   In the evaluation of online success of museums 
\cite{finnis}, and the Social Media Metric for US Federal Agencies \cite{howto}, the 
emphasis is not on evaluating social media efforts for marketing purposes, but to 
provide organisations with tools to be able to understand if their efforts in 
engaging citizens have been successful and, crucially, what defines success. It is 
specifically the emphasis on engagement and collaboration with citizens that makes 
these approaches different from the marketing strategies, which are more focused on 
connecting companies with their clients.

Of direct relevance to our second case study is our previous work, as follows.  
We \cite{pianosi2013} used tools that are freely available 
on-line to analyse social media traffic.   The most basic form of effectiveness 
thus becomes creating social 
media conversation. This includes attracting more and new people, engaging them in 
different actions, and assessing how they participate in conversations both theme-wise 
and among themselves.  Four main measurement approaches were used: (1) growth of 
community, (2) engagement (e.g.\ retweets), (3) content indicators, and 
(4) conversations (e.g.\ number of these). For each category different metrics were 
defined and compared, using, as we have noted, publicly available software tools. 


We \cite{pianosi2013} concluded: ``... although useful in understanding 
the effectiveness of a communication campaign in its numerical terms, the proposed 
methodology can only be the first step of a more in-depth investigation about what 
people can learn during their on-line participation, and what is the perceived impact 
of the process on them, behaviour- or citizenship-wise. Consequently a more in-depth 
analysis of the characteristic of the community and a content analysis of on-line 
conversations is necessary...''.  In this present work we are primarily focused 
on the content analysis of on-line (Twitter) conversations.  
We seek to analyse the semantics of the discourse in a data-driven way.   
The following is concluded by \cite{pianosi2013}: ``top-down communication campaigns 
both predominate and 
are advised by those involved in `social marketing' ... . However, this rarely 
manifests itself through measurable behaviour change ...''.   Thus our approach is, 
in its point of departure and vantage point, bottom-up.  I.e.\ our approach is based
on the observable data.  

Mediated by the latent semantic mapping of the discourse, we will develop semantic 
distance measures between deliberative actions and the aggregate social effect.   
We let the data speak (a Benz\'ecri quotation, noted in \cite{greenacre})
in regard to influence, impact and reach.   

\subsection{Lexical and Linguistic Data from Twitter Narratives}


Twitter data presents all sorts of problems for word, or certainly more so for linguistic,
analysis.  One example from the Stephen Fry tweets that we use is the following that 
is part of a tweet: ``Too twired to teet, too mailed out to e-shag.''  (The first part 
of this play on words and language is referring to ``too tired to tweet'', and the 
second part has even more play on words relating to ``e-mail'' and the informal 
expression for being very tired, ``to be shagged'').  
Another example is a mention of the city of Manchester as ``Madchester'' with its 
``Reet pleased (note stunningly accurate Mancunian accent)''.   
There are many further examples of informal expressions, ``gr8ly'' meaning 
``greatly'', 
and some use of languages other than English (an exchange in Dutch, 
culinary terms in French.) 

One result of our work is to show how semantic properties of words extracted from Twitter
are usable in practical, application-oriented analysis, that is lexically-based, and 
that has potential for revealing the underlying or latent semantics.   
Since our analysis takes into account all pairwise relationships, 
specified through shared associations, therefore there is incorporation of context of 
words and their use.  From the comprehensive set of relationships between tweets, 
between words, and between tweets and words, we have the basis for analysis of 
semantics.  

The methodology used is based on a latent semantic, metric space embedding followed, 
if desired, by induction of a hierarchical clustering, also expressed as the inducing 
of an ultrametric or rooted tree topology.   

\subsection{Two Case Sudies of Twitter Narratives in This Work}

In our first case study, we take impactful tweets and study their role in the Twitter 
narrative.  In the second case study, we take Twitter data from a carefully planned 
campaign to influence through Twitter the environmentally-conscious attitudes as 
manifested in the Twitter medium.  

\section{First Case Study: Impactful Tweets and Their Role in The Twitter Narrative}

We apply the approach used in \cite{becue} to take the text of a narrative and divide it 
into lexically homogeneous subsequences or parts, and coupled with this, to detect 
natural 
breakpoints in the narrative flow.  The advantage of the geometric data analysis approach 
used in \cite{becue} is that the structure of the narrative, and the semantic flow, are 
revealed in a bottom-up manner, based on the actual textual data.  

Our geometrical data analysis approach uses Corresponence Analysis in order to map out 
``the flow of thought and the flow of language'' \cite{chafe1,chafe2} in a 
Euclidean metric, latent semantic factor space.  From that factor space, a hierarchical 
topology is determined, and this hierarchy express the semantics at a continuum of 
resolutions or scales.   

Unlike previous work that uses geometric data analysis on textually-expressed
narrative, including \cite{becue,murtaghganzmckie,murtaghganzreddington}, here in 
this work 
we are involved with social media data where the narrative is much more diffuse and 
less focused.  
Twitter, consisting of streams of text messages called tweets that are each a maximum 
of 140 
characters in length, is very often a dialogue with other tweeters, with names 
preceded by the 
``at'' sign, @, and frequently there is reference to topics that are made linkable 
through being
preceded by the hash 
symbol, $\#$.  In our work here, we use a single tweeter's flow of tweets.  Firstly, 
we map out a
narrative from an individual tweeter's vantage point.  Secondly, we develop new 
approaches for 
assessing effect, and impact if the effect is successful, from within the tweeter's 
narrative flow.

\subsection{A Shock Occurrence in a Social Media Narrative: Narrative Context of
This Event}
\label{shock}

When in October 2009, the actor, presenter and celebrity Stephen Fry announced 
his retirement from Twitter to his near 1 million followers, it was a newsworthy 
event.  It was reported \cite{quinn} that  
``Fry's disagreement with another tweeter began when the latter said `I admire 
and adore' Fry, but that he found his tweets `a bit... boring... (sorry 
Stephen)'.

The tweeter, who said that he had been blocked from viewing Fry's Twitter 
feed, later apologised and acknowledged that Fry suffers from bipolar disorder.''

Having caused major impact among his followers and wider afield, Fry actually 
returned to Twitter, nearly immediately, having had an apology from the offending
tweeter, {\tt @brumplum}. 

The two crucial tweets of Stephen Fry's were as follows.  (In discussion below, we 
refer to them as, respectively, the ``I retire'' tweet or the ``aggression'' tweet.)  

\medskip

6:09 a.m. on 31 October 2009: \\
{\tt @brumplum You've convinced me. I'm obviously not good enough. I \\
retire from Twitter henceforward. Bye everyone.}

\medskip

6.13 a.m. on 31 October 2009: \\
{\tt Think I may have to give up on Twitter. Too much aggression and \\
unkindness around. Pity. Well, it's been fun.}

\medskip

In order to look at those decisive tweets in context, we took a set of 302 of Fry's 
tweets, spanning the critical early morning of 31 October 2009.   These were from 
22 October 2009 to 22 November 2009.  

\subsection{Analysis}


Words are collected from the 302 tweets.  Initially we have 1787 unique words defined as 
follows: containing more than one consecutive letter; with punctuation and special 
characters 
deleted (hence with modification of short URLs, hashtags or Twitter names preceded 
by an at sign,
but, for our purposes not detracting from our interest in semantic content); and with 
no lemmatization
nor other processing, in order to leave us with all available 
emotially-laden or emotionally-indicative function words.  For our 
analysis we do require a certain amount of sharing of words by the tweets.  Otherwise 
there will be 
isolated tweets (that are disconnected through no shared terms).  So we select words  
depending on two thresholds: a minimum global frequency and a minimum number of tweets 
for which 
the word is used.  Both thresholds were set to 5 (determined as a compromise between a 
good overlap 
between tweets in terms of word presence, yet not removing an overly large number of 
words).  This 
led to 143 words retained for the 302 tweets.  A repercussion is that some tweets 
became empty of words: 293 were non-empty, out of the 302 tweet set.    

%

For high dimensional word usage spaces, it is normal for Correspondence Analysis 
to have a lack of concentration of inertia in the succession of factors (cf.\ Appendix 
B), that is to say, the latent semantic factors are of relatively similar importance.
Hence, we developed an analysis methodology as outlined in the following sections.  

\subsubsection{Exploring the Two Critical Tweets in Terms of Their Words}

First we pursued the following analysis approach.  Taking 
the two crucial tweets noted in section \ref{shock}, there were 33 words, as follows.

``to'', ``and'', ``it'', ``on'', ``you'', ``me'', ``not'', ``have'', ``up'', ``too'',
``from'', ``good'', ``well'', ``think'', ``ve'', ``been'', ``may'', ``much'', 
``twitter'', ``fun'', ``brumplum'', ``enough'',``everyone'', ``give'', ``obviously'',
``aggression'', ``around'', ``bye'', ``convinced'', ``henceforward'', ``pity'',``retire'',
``unkindness''  

(Note ``ve'', from ``have'', due to the removal of an apostrophe.)
Then we seek out all other tweets that use at least one of these words.  That resulted 
in 225 out of the total of 302 tweets being retained.

\begin{figure}
\includegraphics[width=12cm]{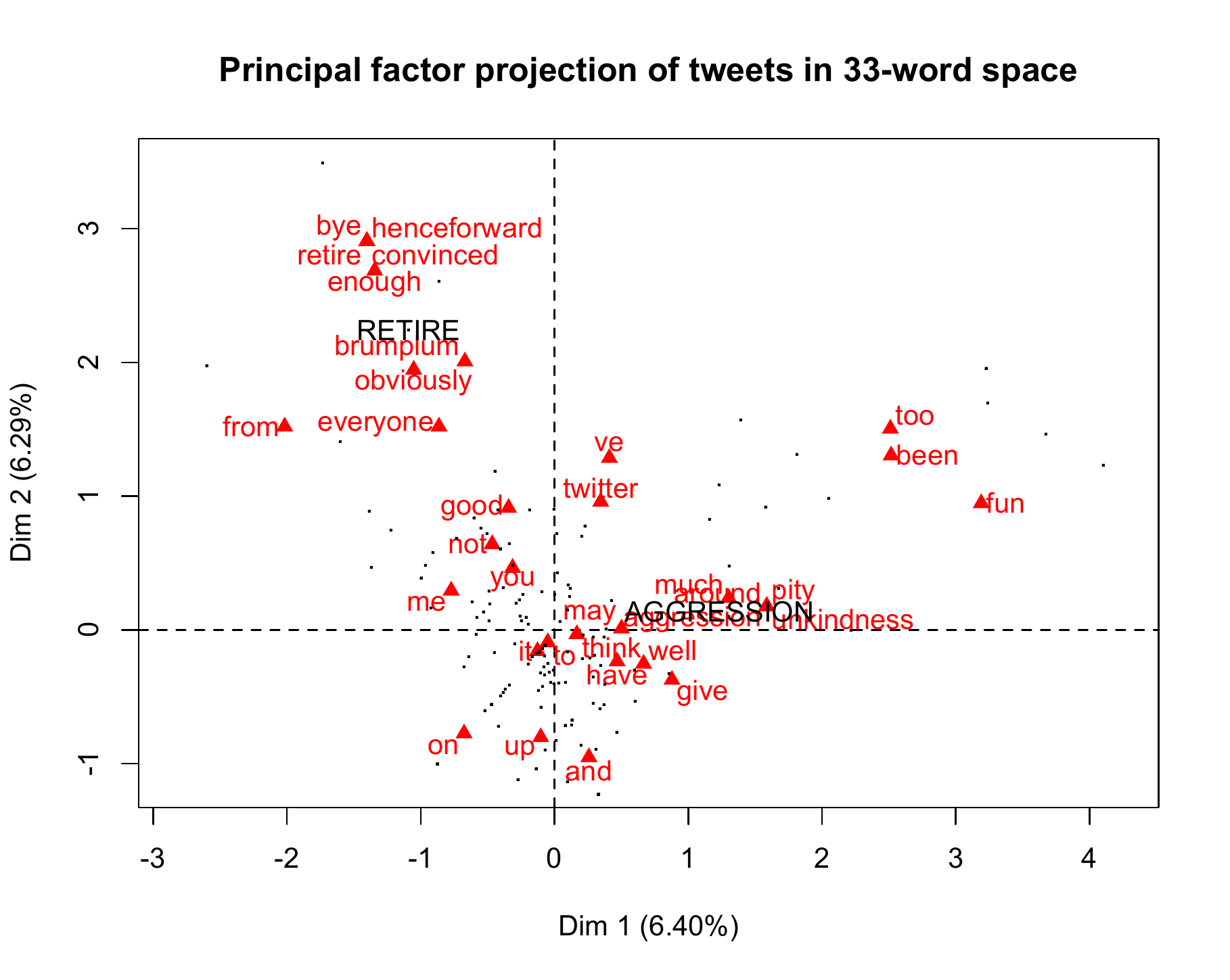}
\caption{Factors 1 and 2, the best two-dimensional or planar projection of the data 
cloud of 302 tweets, where 225 tweets were retained as non-empty.  Simultaneously we 
have the planar projection of the 33 word cloud. The dots are at the locations of the 
tweets (identifiers are not shown, to avoid overcrowding).   Just two tweets, the 
crucial two, have the ``retire'' and ``aggression'' labels (and not just a dot).}  
\label{Fry1}
\end{figure}

Figure \ref{Fry1} positions our two critical tweets in a best planar projection of the 
tweets and associated words.  In Figure \ref{Fry1}, the contribution to the inertia of 
factor 1 by the ``aggression'' tweet is the greatest among all tweets, and the 
contribution to the inertia of factor 2 by the ``I retire'' tweet is the greatest 
among all tweets. 
 While useful for finding dominant themes (expressed by the words used in the tweets), 
and perhaps also for the trajectory of these themes, we can use the full dimenensionality
of the latent semantic representation of this Twitter data by clustering the tweets, based
on their (Euclidean metric) factor projections.   
We use a chronologically (or sequence) constrained complete link agglomerative 
hierarchical clustering.  
See \cite{legendre,murtaghganzmckie,becue} for this hierarchical clustering approach.

\begin{figure}
\includegraphics[width=12cm]{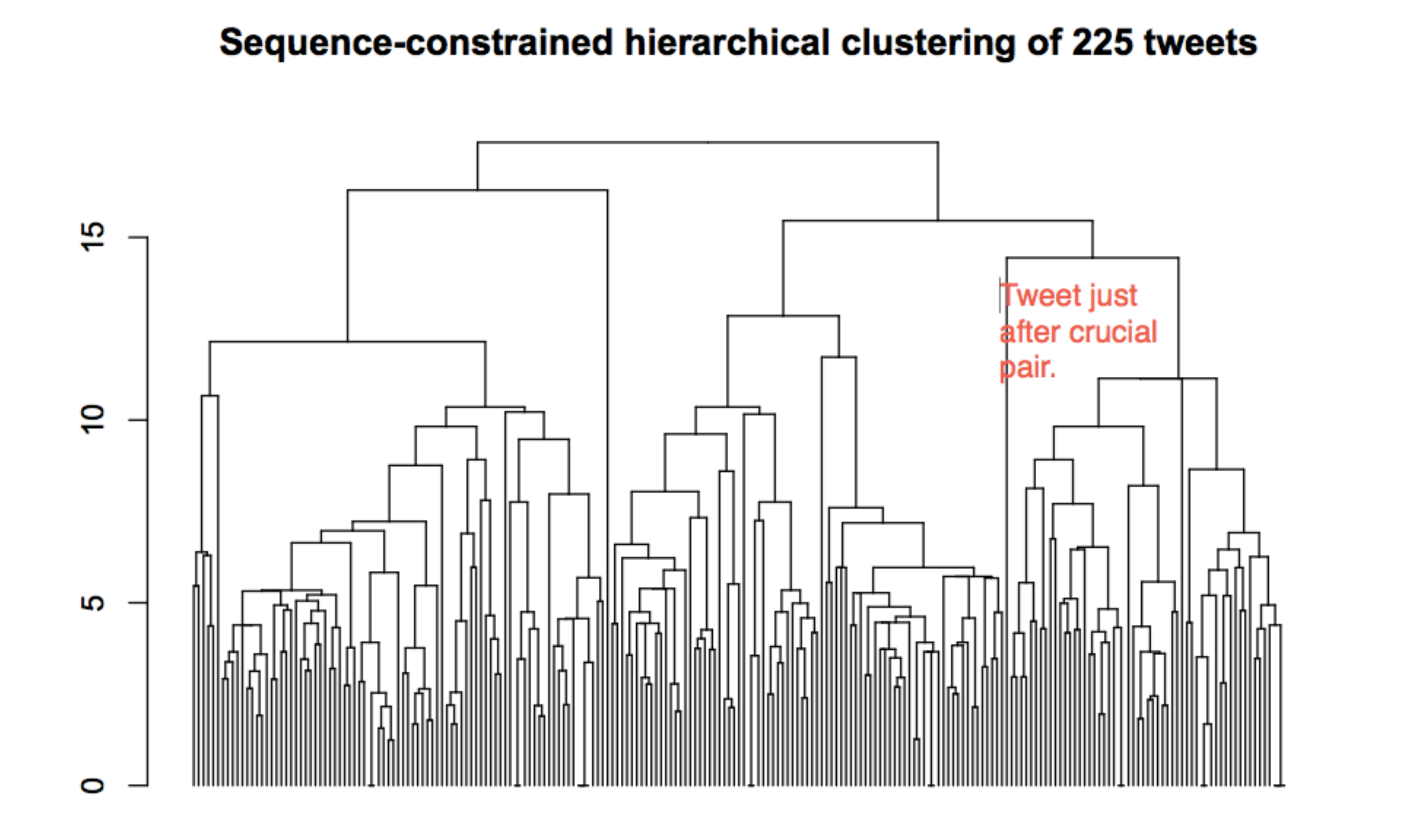}
\caption{Hierarchical clustering, using the complete link agglomerative criterion
(good for compact clusters) on the full dimensionality, Euclidean factor coordinates.
Just 33 words are used.
The tweet (with a relatively long branching path before it is agglomerated, to the left 
side of the text) is annotated that is 
immediately following the two crucial ones that we are focused on, i.e.
the ``I retire'' tweet and the ``aggression'' tweet.}
\label{Fry2}
\end{figure}

\begin{figure}
\includegraphics[width=12cm]{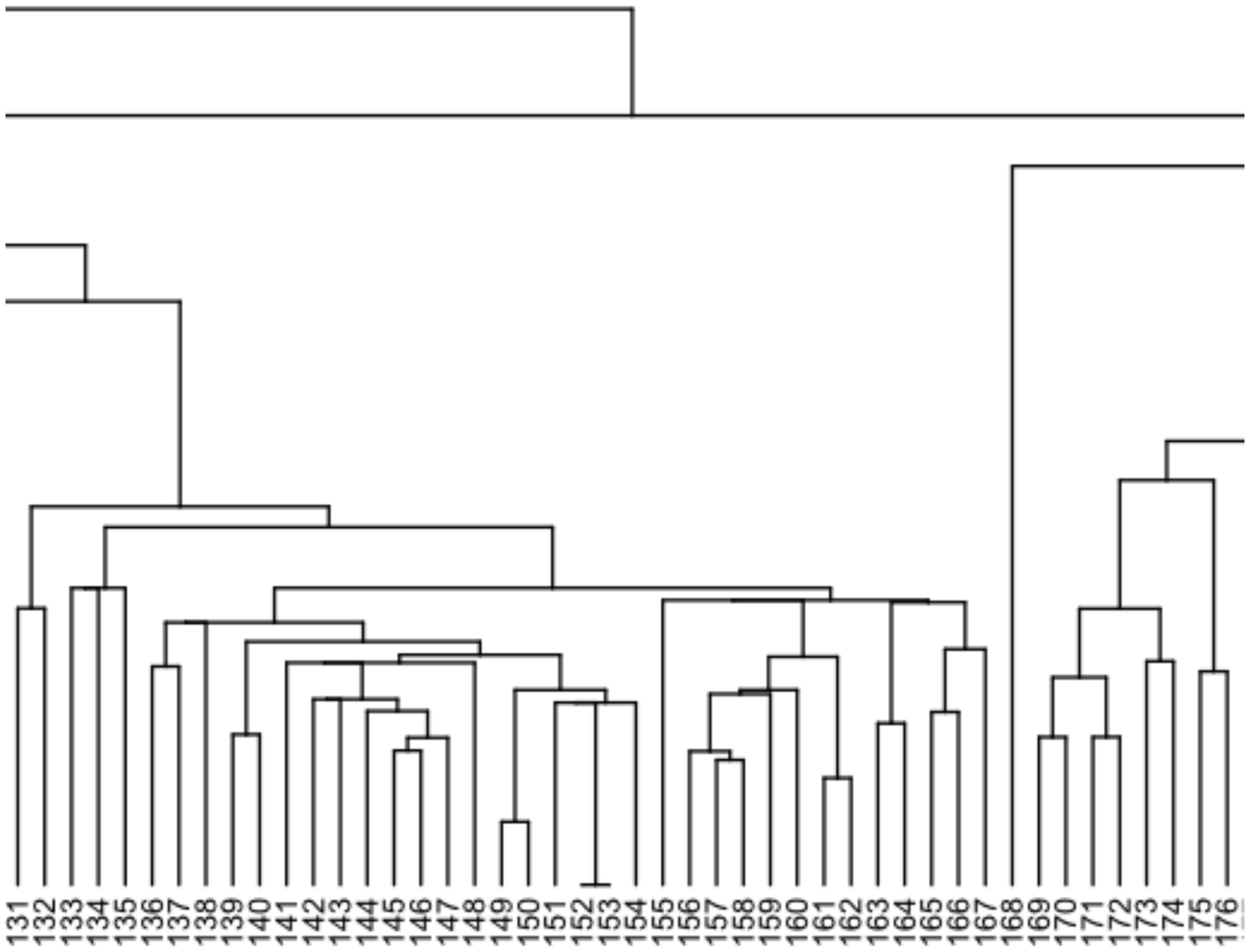}
\caption{A close-up from Figure \ref{Fry2}.  Our two critical tweets
are the 166th and 167th ones here, the ``I retire'' and ``aggression'' tweets.
(Cf.\ section \ref{shock}.)}
\label{Fry2b}
\end{figure}

Figure \ref{Fry2} displays this hierarchical clustering of the Twitter narrative.  
Figure \ref{Fry2b} is a close-up view of part of the dendrogram.   Our crucial tweets 
are located at the end of a fairly compact clustering structure.  This points to how 
our two crucial tweets can be considered as bringing a sub-narrative to a conclusion.  
Our interest is therefore raised in finding sub-narratives in the Twitter flow.  These 
sub-narratives are sought here as chronologically 
successive tweets, i.e.\ a segment in the chronological flow of tweets.  

%
%
%
%
%
%
%
%
%
%
%
%

\subsubsection{Exploring the Two Critical Tweets in Terms of Twitter Sub-Narratives}

To investigate our two critical tweets, such as the immediate or other precursors,
and the repercussions or subsequent evolution of the Twitter narrative, 
we will now determine sub-narratives in the overall Twitter narrative.  This we will do 
through segmentation of the flow of tweets.  So a sub-narrative is defined as a segment of 
this flow of tweets.  That is, the sub-narrative consists of groups (or clusters) of successive 
tweets that are semantically homogeneous.  Semantic homogeneity is defined through a statistical 
significance approach.  

\subsubsection{Sub-Narratives, Twitter Data Used, Hierarchical Structure of the 
Overall Twitter Narrative}

We return now to the full, original word set. 

On the full set of tweets and the words used in these tweets, a threshold of 5 tweets was 
required for each word, and the total number of occurrences of words needed to be at least 5.  
This lowered our word set, initially 1787, to 143.  Then we removed stopwords, and partial words, 
 in a list that 
we made: ``the'', ``
to'', ``and'', ``of'', ``in.'', ``it'', ``is'', ``for.'', ``that'', ``on'', ``at'',
``be'', ``this'', ``what'', ``an'', ``if.'', ``ve'', ``don'', ``ly'', ``th'', ``tr'', ``ll''.
That led to 121 words retained.  There remained
280 non-empty tweets (from the inital set of 302 tweets).  Our two critical tweets (the ``I retire''
and the ``aggression'' ones) were among the retained tweet set.  

\begin{figure}
\includegraphics[width=12cm]{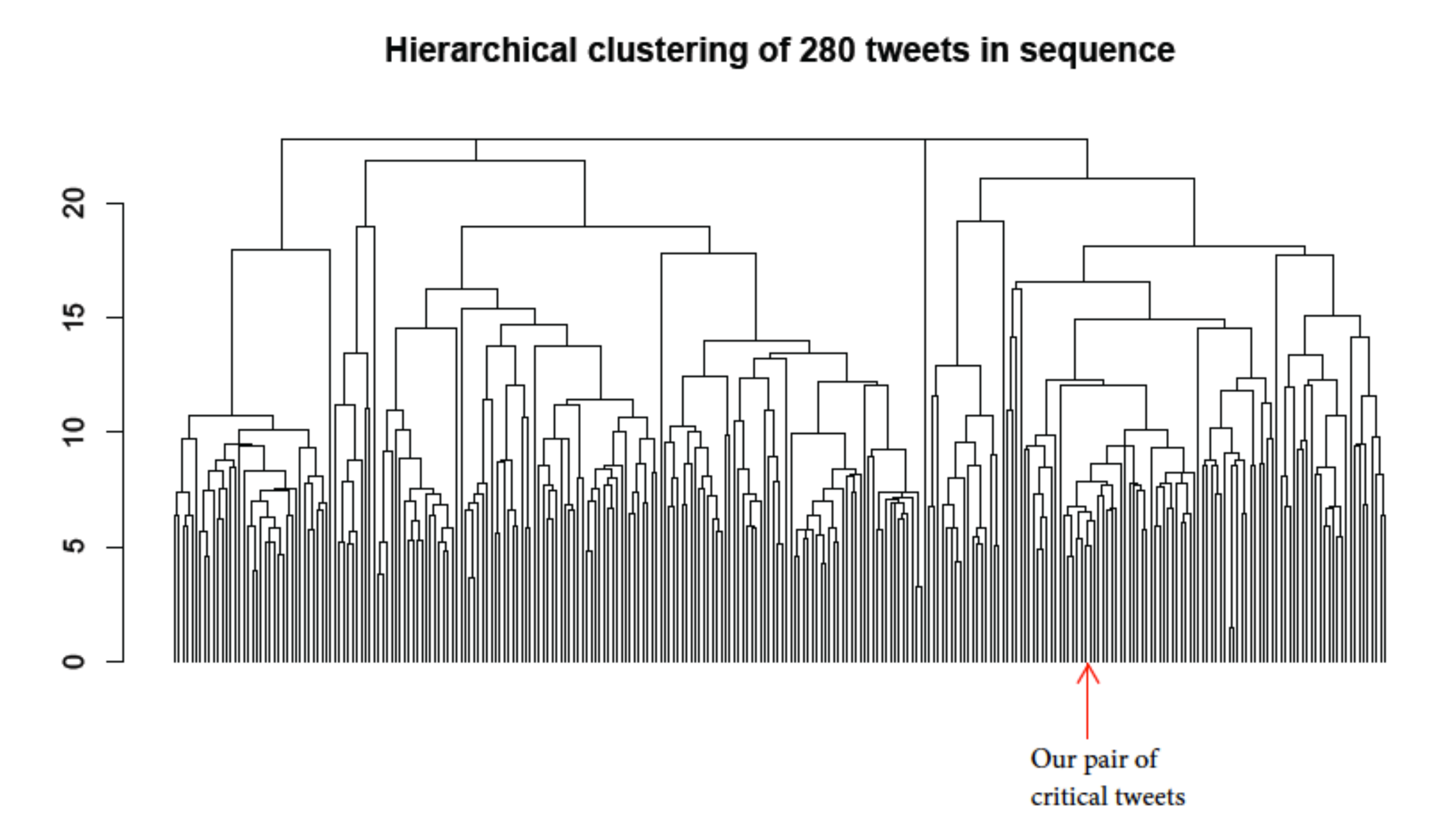}
\caption{Hierarchical clustering, using the complete link agglomerative criterion
(providing compact clusters) on the full dimensionality, Euclidean factor coordinates.  
The tweets are characterized by presence of any of the 121 word set used. The 
280 tweets, in chronological sequence are associated with the terminal nodes (arranged 
horizontally at the bottom of the dendrogram or hierarchical tree). 
We look for an understanding of semantic content, and the evolution of this, leading up
to our two crucial tweets, and the further evolution of the tweet flow.}
\label{Fry3}
\end{figure}

Following Correspondence Analysis of the 280 tweets crossed by 121 words, 
an agglomerative hierarchical clustering was applied on the full-dimensionality
factor space coordinates. 
The chronological sequence of tweets was hierarchically clustered.   
With the set of 280 tweets, crossed by the 121 word set, Figure \ref{Fry3} shows the chronological
hierarchical clustering.  Our two critical tweets are in their chronological sequence in the 
280-tweet sequence (at the 211th and 212th tweet positions in this sequence).  

A note follows now on why we did not use hashtag words (themes referred to), or at-sign prefaced 
words (other tweeters by Twitter name).   
The hashtag was not used all that often, the usages being:
{\tt \#media140, \#thearchers, \#frys, \#FryS, \#140conf,  
\#grandslamdarts, \#pdc} (previous two generally together), 
{\tt \#webwar, \#threestrikes} (previous two together always), {\tt \#svuk}.  The total number of 
at-sign names was 86.  This was insufficient to base our entire analysis on Twitter names, even 
if hashtag themes were added.   

To exploit the visualization of the Twitter narrative that is expressed in
Figure \ref{Fry3}, we will summarize this visualization by determining a segmentation of the 
flow of tweets.  That is equivalently expressed as determining a partition of tweets 
from the dendrogram.  Furthermore, as described in the next subsection, we look for internal 
nodes of the dendrogram that are statistically significant (using the approach that will now 
be described).

\subsubsection{Sub-Narratives of the Overall Twitter Narrative through Segmenting the Twitter 
Flow}

In line with \cite{becue}, we made these agglomerations subject to a permutation test to 
authorize or not each agglomeration that is deemed to be significant.  
In the description that now follows for determining significant segments of tweets, we 
follow very closely \cite{becue}.  Statistical significance is that the agglomerands 
validly form a single segment.   

All the distances between pairs of objects of the two adjacent groups that are candidates
for agglomeration are computed. These distances are divided into two groups: 50\% of the 
distances with the highest values are coded with 1 and 50\% with the lowest values are 
coded with 0. The count of high distances 
is denoted by $h$.  The count of high distances between permuted groups is also computed.
 
The number of permutations producing a result equal to or over $h$, divided by the
number of permutations that are performed, gives an estimate of the probability $p$ of
observing the data under the null hypothesis (the objects in the two groups are drawn
from the same statistical population and, consequently, it is only an artefact of the
agglomerative clustering algorithm that they temporarily form two groups).
Probability $p$ is compared with a pre-established significance level $\alpha$. 
If $p > \alpha$, the null hypothesis is accepted and the fusion of the two groups is 
carried out. If $p \leq \alpha$, the null hypothesis is rejected and fusion of the 
groups is prevented.  Changing the value of $\alpha$ changes the resolution of the 
partition obtained, which is what is obtained when the sequence of agglomerations is 
not allowed to go to its culmination point (of just one cluster containing all entities 
being clustered).   

An $\alpha$ significance level of 0.15 was set (giving an intermediate number of segments 
between not too large, if $\alpha$ were set to a greater value, or a small number of segments
if the significance level were more demanding, i.e.\ smaller in value).  Assessment of 
significance used 5000 permutations (found to be very stable relative to a number of permutations 
that were a few hundred upwards).  

The number of segments found was 40.  
A factor space mapping of these 40 segments was determined, in their 121-word space. Four of these
segments (6th, 18th, 36th, 39th) had just one tweet.  Since they would therefore quite possibly 
perturb the Correspondence Analysis, in being exceptional in this way, we took these particular
tweets as supplementary tweets.  This means that the Correspondence Analysis factor space 
(i.e.\ the latent semantic space endowed with the Euclidean metric) was determined using the 
active set of 40 less these four tweets, and then the four supplementary tweets were projected into
the factor space.  

\begin{figure}
\includegraphics[width=12cm]{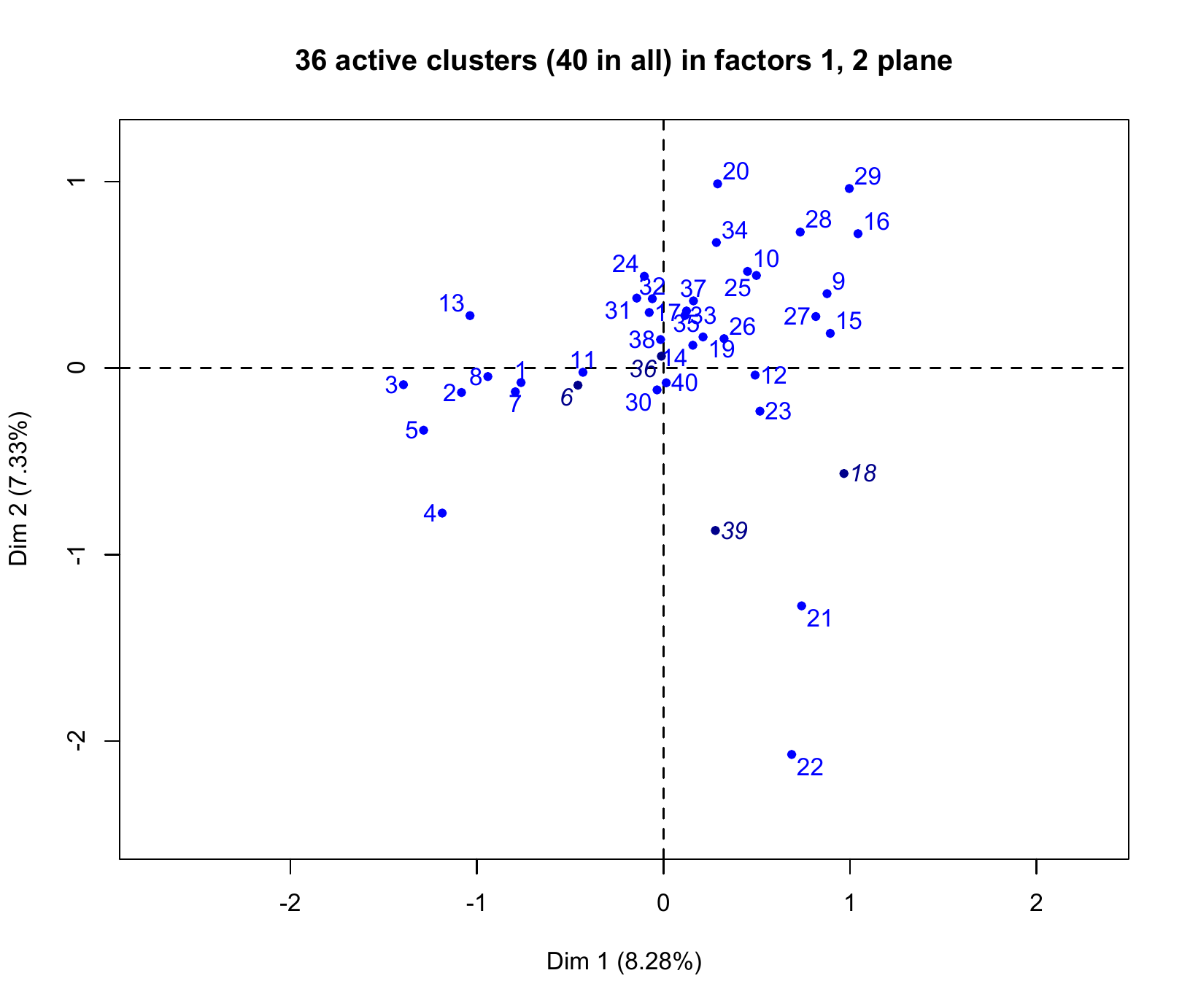}
\caption{The centres of gravity of 40 segment groups of the Twitter flow are projected in 
the principal factor plane.  (See text for details related to 36 of these tweets being used
for this analysis, and then 4 being projected into the factor space as supplementary tweets.)}
\label{Fry4}
\end{figure}

The mapping is shown in Figure \ref{Fry4}.
It is noticeable that segment group 30, that contains our critical tweets towards the end of it,
is very close to the origin, which is the average tweet here.  The average tweet can be 
taken as the most innocuous.  Therefore the factor plane of factors 1 and 2 is not useful for
saying anything further about segment group 30, beyond the fact that it is fully unremarkable.

The contributions of segment group 30 to the factors 1, 2, 3, 4, 5 are, respectively 0.04, 
1.71, 9.94, 1.62, 0.27.  We will look at factors 2,3 because they are determined far more 
(than the other factors here) by segment group 30. 

\begin{figure}
\includegraphics[width=12cm]{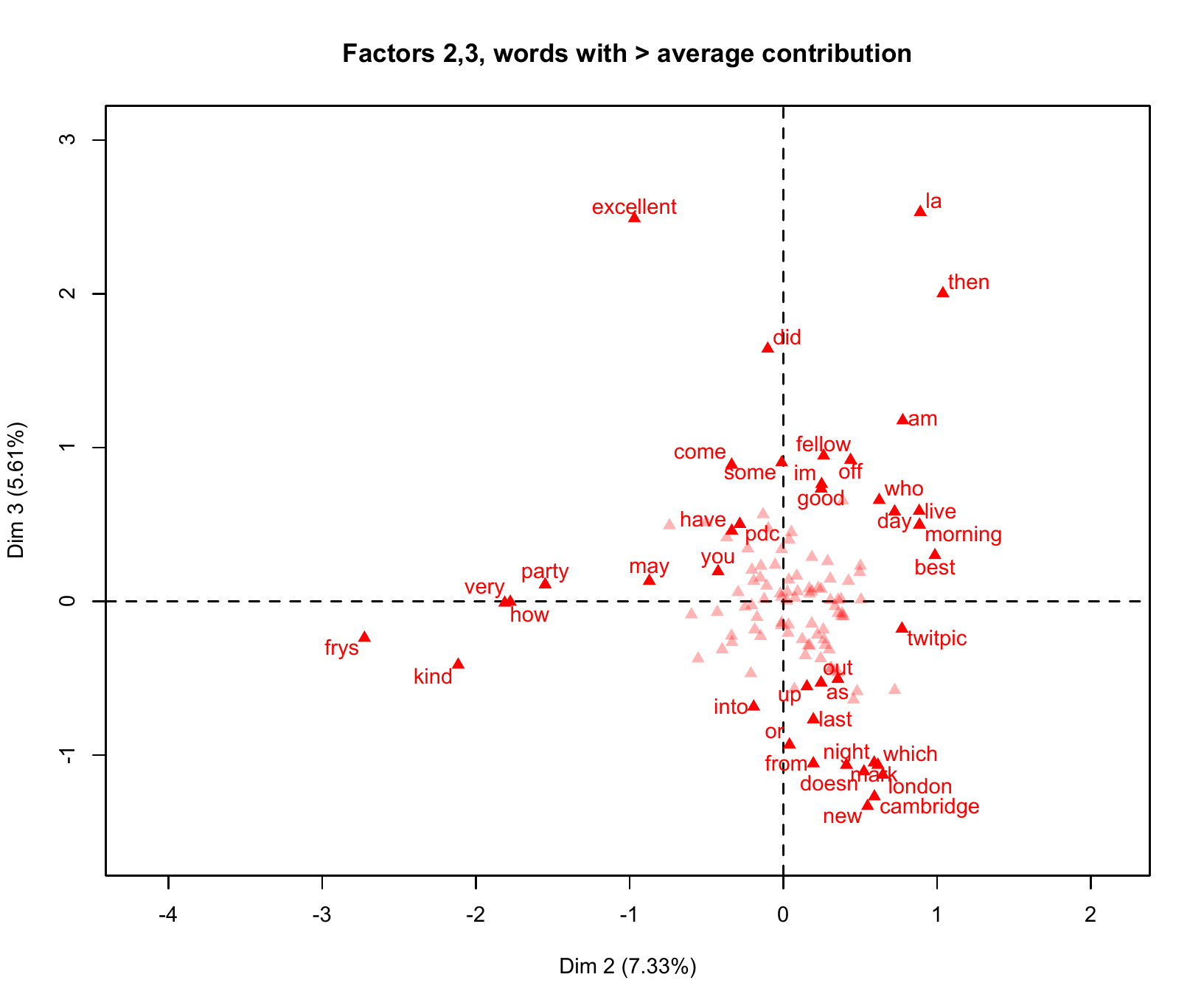}
\caption{Important words, with contribution to the inertia of the cloud of all words in this factor
plane, of factors 2,3.}
\label{Fry6}
\end{figure}

\begin{figure}
\includegraphics[width=12cm]{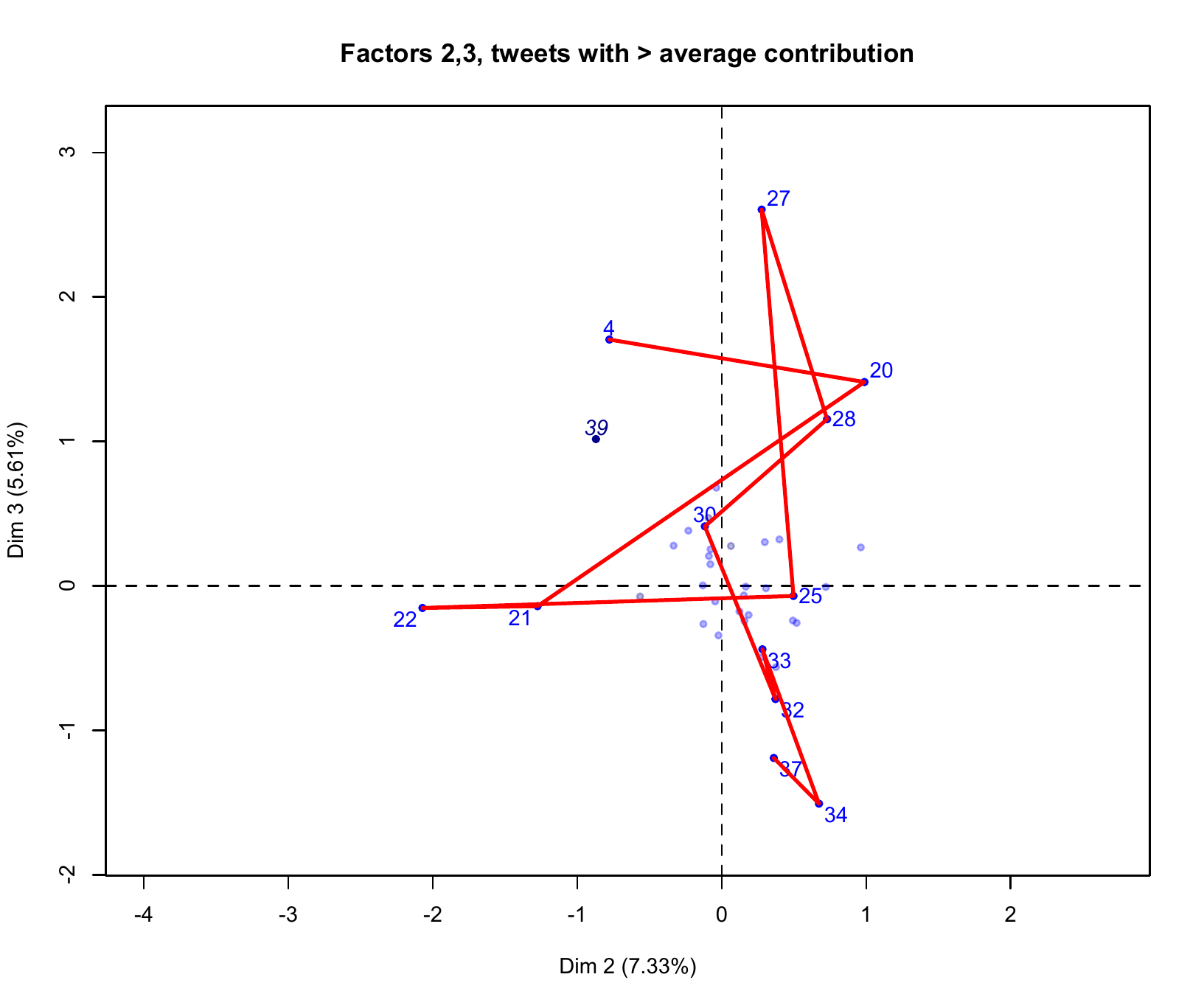}
\caption{The plane of factors 2,3 with the important tweet segments.  These 
tweet segments are important 
due to greater than mean contribution to 
the inertia of the cloud of tweet segments.  The trajectories connecting the tweets
in their chronological order are also shown.}
\label{Fry7}
\end{figure}

Figure \ref{Fry6} displays the words that are of greater contribution to the mean inertia of 
factors 2 and 3.  We note that
Figure \ref{Fry7} displays the important tweet segments 
in the factors 2,3 plane, i.e.\ the tweet segments with 
contribution to the inertias of these factors that are greater than the mean.  The 
chronological trajectory linking these important tweet segments is also shown.  

Early tweet segments are positive on factor 3.  Then there is a phase (with important 
tweet segments  21, 22, 25) that are fairly neutral on factor 3, but range, first negatively
on factor 2, and then positively.  Then comes a phase (through tweet segment 27) 
of strong factor 3 positivity.  Recall that positive and negative orientations of factor axes
are relative only and contain no judgemental character whatsoever.   With tweet segment 28, there
is a move that is reinforced by tweet segment 30, containing our crucial two tweets, back towards
the other extremity of factor 3.  Further tweet segments then play out their roles on the 
negative factor 3 half-axis.  

In summary we find the following description of the segment groups.

Positive factor 3: segment group 27, appearance in LA (Los Angeles); segment group 28, relating 
to appearance on the morning news and entertainment television 
show ``Good Day, LA''; segment group 20, recording of British science fiction television series, 
``Dr.\ Who''.

Negative factor 3: segment group 34, Cambridge (England) and (London Street) Norwich; segment 
group 37, London (England).   Segment group 32 concerns computer-related purchases and issues, and 
a London event; segment group 33 relates to Royal Geographic Society and other events.

So our tweet segment of interest, segment group 30, is between tweets that are 
mainly dealing with LA and the London area. In segment 
group 30, there is the alternation with Twitter user {\tt @brumplum}, 
and also a mention of having arrived
in LA.  We note therefore these geographic linkages in the Twitter vicinity of 
the crucial tweets relating to ``aggression'' and ``I retire''.  Furthermore we note the transition 
back to the London area, where events that Stephen Fry was involved in were based.  

\subsubsection{Conclusion on Study of Impactful Tweets in Overall Narrative}

The completes our analysis of the Stephen Fry case study.  We have described initally
how we did not find anything remarkable in the narrative flow relating to our two 
crucial tweets.  Then we pursued analysis of sub-narratives, determined by segments 
of the narrative flow.  We found a number of special characteristics both of, and 
closely related to, the two crucial tweets.

\section{Impact and Effect in Twitter Narrative Relating to Environmental Citizenship}

Our next case study relates to the furthering and encouragement of environmental citizenship, 
i.e.\ engagement and responsibility in regard to environmental issues.   
The background to this work encompasses the following aspects: (1) the testing of 
social media 
with the aim of designing interventions; (2) application to environmental 
communication initiatives; and (3) measuring impact of public engagement theory.  The 
latter aspect is in the (renowned social and 
political science theorist) J\"urgen Habermas sense of public engagement centred on communicative 
theory. By implication therefore, this points to discourse
as a possible route to social learning and environmental citizenship.  For us, here, 
discourse is Twitter-based.  

In \cite{pianosi2013} we deal with the practical challenge of how on-line 
activity can actually be measured.  The Twitter campaigns set up and used in  
\cite{pianosi2013} -- which we 
use in this work -- are considered as discourse-based media and as such they have links 
with public engagement centred on Habermas's communicative theory.  In order to define 
and then measure terms like 
``influence'', ``impact'' and ``reach'', we sought, in 
\cite{pianosi2013}, to evaluate if this is simply the number of friends, followers, 
re-tweets or ``like'' in a social media (Twitter, Facebook) setting, and whether such 
social media actions could be considered, in appropriate circumstances, as an act of 
citizenship or public engagement.

\subsection{Input Data Structure, Mapping into a Semantic Space}

\subsubsection{Innovation in This Work: How We Address Impact}

For us, {\em Impact} will be the semantic distance between the initiating action, 
and the net aggregate outcome.  
This can be statistically tested.   It can be visualized.  
Facets and indeed components of such impact can be further visualized and evaluated.  

Essential enabling aspects are (1) the {\em data structure input}, 
comprising characterization 
of relevant actions, characterization of the initiating actions; and for all relevant 
actions, and the initiating actions, we have their context mode (called 
``campaign'' here) which allows both intra and inter analyses.  (2) {\em Mapping} of this 
characterization data (presence/absence, frequency of occurrence, mode category) 
{\em into 
a semantic space} that is both qualitatively (through visualization) and quantitatively 
analyzed.  This semantic space is a Euclidean, factor space.  

For visualization we use 
2-dimensional projection, but for quantitative analysis, we use the full factor space 
dimensionality, hence with no loss of information.    


\subsubsection{The Data Used}

The eight campaigns in late 2012 were as follows, with the date during which the 
campaign was carried out, and the theme of the campaign. 

\begin{enumerate}
\item 1 October to 7 October: Climate change: The big picture and the global 
consequences.
\item 8 October to 14 October: Climate change: The local consequences.
\item 15 October to 22 October: Light and electricity.
\item 23 October to 28 October: Heating systems.
\item 29 October to 4 November: Sustainable Food choices.
\item 5 November to 11 November: Sustainable Travel choices.
\item 12 November to 18 November: Sustainable Water use.
\item 19 November to 25 November: Sustainable Waste.
\end{enumerate}

\begin{table}
\begin{center}
\begin{tabular}{|llll|}\hline
Seq. no. & Tweet & Init. -- yes/no & Campaign 1, 2, ..., 8 \\ \hline
1 & Tweet 1 & 1  &  1  \\
2 & Tweet 2 & 0  &  1  \\
... & ... & ... & ... \\
... & ... & ... & ... \\
985 & Tweet 985 & 0 & 8 \\ \hline
\end{tabular}
\end{center}
\caption{Transformed Twitter data used.  Column 1 is the sequence number of the tweet.
Column 2 is the tweet. Column 3 has the value 1 if the tweet was an initiating one for 
a new campaign, and otherwise is 0.  Column 4 has the value 1 to 8, indicating
the campaign.}
\label{tab1}
\end{table}

Table \ref{tab1} depicts the initial data set derived from the Twitter data spanning 
the eight campaigns.   There are 985 tweets here.   
Campaigns were as follows in the succession of tweets: 1 to 63; 64 to 133; 134 to 
301; 302 to 409; 410 to 555; 556 to 730; 731 to 843; and 844 to 985.  
The initiating tweets for the eight campaigns are: 3, 65, 134, 303 and 304 (which 
were combined -- the two taken together as one), 410, 557, 736 and 846.  These 
initiating tweets are listed in full in Appendix A.  

In the first stage of the processing, from all tweets a set of
3056 terms was derived.  These terms were essentially 
the full word set obtained from all tweets.  See below, in the following subsection,
for an exact specification. 
Each tweet was cross-tabulated with those terms that were present for 
it.  (Storage-wise, each tweet had 1 = presence, 0 = absence values for each of the 
3056 terms.  In some cases there were 2 or 3 presences.)
In a second stage of the processing, the term set was reduced to 339 sufficiently 
often used terms.   Some tweets thereby 
became empty, so the number of usable, non-empty tweets dropped from 985 to 968
non-initiating tweets plus the 8 initiating tweets.  (We have already noted that seven 
of the eight campaigns had one initiating tweet.  Campaign 4 had two 
successive initiating tweets.
We joined these two tweets together into a single initiating tweet for campaign 4.)

For the Correspondence Analysis, the latent semantic mapping method used, the
 input data set used is depicted in Table \ref{tab2}.
For the analysis, we distinguish between principal rows (tweets that are not 
initiating ones) and supplementary rows (tweets that are initiating ones); 
and principal columns (terms used by the tweets) and supplementary columns
(categorization in regard to the campaign).  See Table \ref{tab2}.   The analysis
that embeds rows and columns in a factor space is carried out on the principal rows
and columns, i.e.\ the regular discourse (non-initiating) tweets crossed by the 
terms that characterize them.  Into that factor space, the supplementary rows and 
columns are projected, i.e.\ respectively the initiating tweets, and the 
campaign categories.  

The data to be analyzed then was as follows.

\begin{itemize}
\item Principal rows: the set of 968 retained tweets, that do not include the 
initiating tweets.  
\item Supplementary rows: the set of 8 initiating tweets.  
\item Principal columns: the set of 339 terms retained.  
\item Supplementary columns: the set of 8 ``indicators'' for the 8 campaigns.
\end{itemize}

\begin{table}
\begin{center}
{\large Tweets}
$\left\{
\begin{tabular}{c}
  \\
  \\
  \\
  \\
  \\
  \\
  \\
  \\
  \\
  \\
  \\
  \\
  \\
  \\
\end{tabular}
\right.$
\begin{tabular}{ccccccccccccccccc} 
\multicolumn{15}{c}{$\overbrace{\rule{3cm}{0pt}}^{\mbox{\large Terms}}$} & \multicolumn{2}{c}{$\overbrace{\rule{1cm}{0pt}}^{\mbox{\large Cats.}}$} \\ \hline
\multicolumn{15}{|c}{} & \multicolumn{2}{|c|}{} \\ 
\multicolumn{15}{|c}{} & \multicolumn{2}{|c|}{} \\ 
\multicolumn{15}{|c}{} & \multicolumn{2}{|c|}{} \\ 
\multicolumn{15}{|c}{} & \multicolumn{2}{|c|}{} \\ 
\multicolumn{15}{|c}{} & \multicolumn{2}{|c|}{} \\ 
\multicolumn{15}{|c}{} & \multicolumn{2}{|c|}{} \\ 
\multicolumn{15}{|c}{} & \multicolumn{2}{|c|}{} \\ 
\multicolumn{15}{|c}{} & \multicolumn{2}{|c|}{} \\ 
\multicolumn{15}{|c}{} & \multicolumn{2}{|c|}{} \\ 
\multicolumn{15}{|c}{} & \multicolumn{2}{|c|}{} \\ 
\multicolumn{15}{|c}{} & \multicolumn{2}{|c|}{} \\ 
\multicolumn{15}{|c}{} & \multicolumn{2}{|c|}{} \\ 
\multicolumn{15}{|c}{} & \multicolumn{2}{|c|}{} \\ 
\multicolumn{15}{|c}{} & \multicolumn{2}{|c|}{} \\ 
\multicolumn{15}{|c}{} & \multicolumn{2}{|c|}{} \\ 
\multicolumn{15}{|c}{} & \multicolumn{2}{|c|}{} \\ \hline
\end{tabular} 
\\
\vspace*{-0.2cm} 
{\large $\overset{\mbox{Init.}}{\mbox{Tweets}}$} 
$\left\{
\begin{tabular}{c}
  \\
  \\
\end{tabular}
\right.$
\begin{tabular}{ccccccccccccccccc} 
\multicolumn{15}{|c}{  \ \ \ \ \ \ \ \ \ \ \  \ \ \ \ \ \ \ \ \ \ \ \ \ \ \ \ } & \multicolumn{2}{|c|}{ \ \ \ \ \ \ \ \ \ } \\  
\multicolumn{15}{|c}{} & \multicolumn{2}{|c|}{} \\ \hline 
\end{tabular}
\end{center}
\caption{Upper left, Tweets $\times$ Terms: very sparse, most values 0 indicating
absence of term in the tweet.  Some values 1 (and a few 2 or even 3) 
indicating presence of term in the tweet.
Upper right, Tweets $\times$ Categories, 1 in the 
relevant campaign column associated with the tweet.  Otherwise 0.  
Lower left, Initiators $\times$ Terms: as for Tweets $\times$ Terms.  
Lower right, Initiators $\times$ Categories, i.e.\ Campaigns: each row has a campaign = 1 
and otherwise 0.}
\label{tab2}
\end{table}

\subsubsection{Preprocessing the Set of Terms, i.e. the Content of the Tweets}
\label{preprocess1}

In this and the next subsection, we explain how we select the term set used to 
characterize each tweet in the overall Twitter discourse.  

Only alphabetic characters are retained.  So 
@, \# are dropped but we can generally spot user or hashtag terms from the 
remaining term stump.  Numerical data are dropped including dates, since we 
will focus exclusively on word-based data.  Punctuation and special characters 
go too, e.g.\ in URLs.  We could handle these were it advisable to do so.

The html expression for ampersand (``\&amp;''), in our processing left with a 
rump, ``amp'', is substituted with ``and''.

All upper case is set to lower case (with no loss of information involved).

We deleted ``ll'' (left over from e.g.\ ``I'll''), ``s'' 
(from e.g. ``it's''), and ``t'' (from e.g.\ 
``isn't'' or ``wouldn't''). 

We find 3323 terms used in the original set of 985 tweets. 

Terms on a stopword list (``and'', ``the'', etc.) are deleted and this decreases the 
3323 term set to 3056 terms.  In \cite{ref08888} (section 5.3, ``Tool words: between
analysis of form and analysis of content'') we discuss the case for considering
such function words or ``tool words'' in many languages, that are especially important
for characterizing style.  However this present work relates to within a single 
discourse.  

\subsubsection{Preprocessing the Tweets $\times$ Terms Matrix}
\label{preprocess2}

The tweets $\times$ terms cross-tabulation is set up, with frequency of 
occurrence values.   The greatest frequency of occurrence value is 3.  
Typically the frequency of occurrence is 1.  The cross-tabulation matrix 
is very sparse, with most values equal to 0.

In order to facilitate and even to make possible the comparison of all tweets 
in the Twitter discourse, 
we require each set of presences of terms over all tweets to be at 
least 5, and also that the term be present in 5 tweets.   Exceptionally rare 
terms would hinder our analysis.  Our thresholds of 5 were such that rarely 
used terms were pinpointed, and not at the cost of removing too many terms.  

The 968 retained (non-initiating) tweets, and the 8 initiating tweets, are 
crossed by 339 terms.  

   



\subsection{Data Analysis}

\subsubsection{Semantic Mapping of Tweets, Terms and Categories through 
Correspondence Analysis}

Factors, in decreasing order of importance, provide latent semantic components.
Analysis is carried out on the principal rows, columns.  Then the supplementary rows, 
columns are projected into the analysis.   The principal rows are the discourse,
non-initiating tweets.  The principal columns are the set of terms used in this 
discourse.  The supplementary rows are the initiating tweets.  The supplementary 
columns are the campaign indicators.  

Factors, in decreasing order of importance, provide latent semantic components.
Analysis is carried out on the principal rows, columns.  Then the supplementary rows, 
columns are projected into the analysis.   The principal rows are the discourse,
non-initiating tweets.  The principal columns are the set of terms used in this 
discourse.  The supplementary rows are the initiating tweets.  The supplementary 
columns are the campaign indicators.  

Each term is at the centre of gravity of ``its'' tweets.  Each tweet is at the centre 
of gravity of ``its'' terms.
The factor space is a semantic space in that it takes account of all interrelationships 
-- between all tweets, between all terms, between all tweets and all terms.  

Typically we visualize this semantic, factor representation of the data by taking two 
factors at a time.  Planar projections lend themselves to such display.  
In the analysis discussion to follow, we tidy up these displays, in order to highlight 
useful and/or important outcomes.

\subsection{Semantically Locating the Initiating Tweets and the 
Net Overall Campaign Tweets}

Our first analysis shows the principal factor plane of the 8 tweets that initiated
the campaigns, where we projected the supplementary rows (cf.\ Table \ref{tab2})
to have their semantic locations; and the net aggregate campaigns, given by the 
centres of gravity of the 8 campaigns, where we projected the supplementary 
columns (cf.\ Table \ref{tab2}) to have their semantic locations.  The actual 
definition of the factors was from the principal rows -- all tweets save the 
initiating ones -- and the principal columns -- the word set used in the Twitter
discourse. 

\begin{figure}
\centering
\includegraphics[width=14cm]{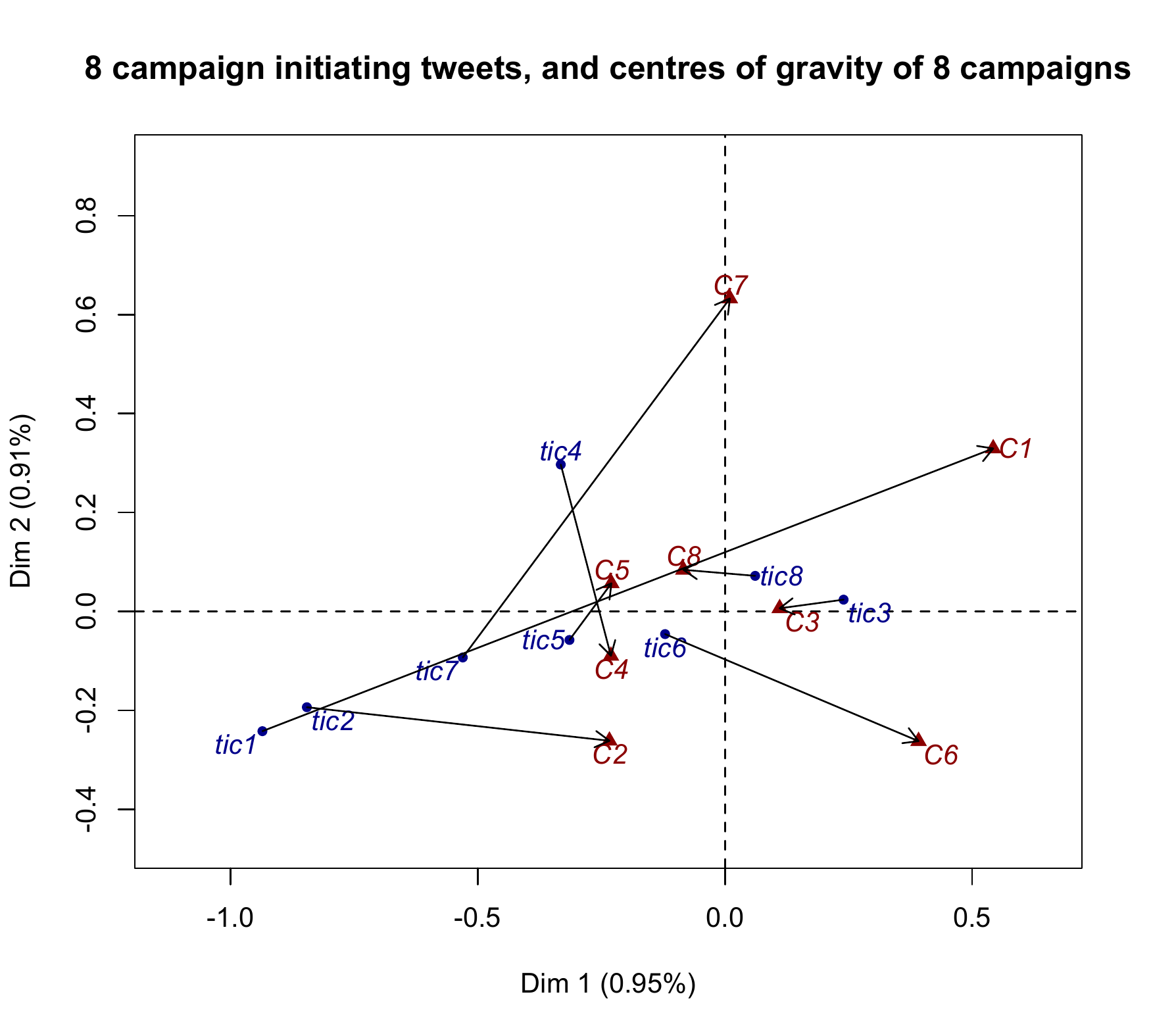}
\caption{The campaign initiating tweets are labelled ``tic1'' to ``tic8''.  
The centres of gravity of the campaigns, i.e.\ the net aggregate of the campaigns, are 
labelled ``C1'' to ``C8''.  In each case, the tweet initiating the campaign is 
linked with an arrow to the net aggregate of the campaign. The percentage inertia
explained by the factors, ``Dim 1'' being factor 1, and ``Dim 2'' being factor 2,
is noted.}
\label{fig1}
\end{figure}

Even if the principal factor plane accounts for relatively little information 
in our data, it nonetheless is the mathematically best planar representation, hence
summary, of our data. 
In this factor 1, factor 2 plane, 
Figure \ref{fig1} shows the instigating tweet (``tic1'', etc.) and the net overall
effect (``C1'', etc.).  

We see that campaigns 3, 5, 8 have initiating tweets that are fairly close to the 
net overall campaign in these cases.  
By looking at all tweets, and all terms, it is seen that the campaign 
initiating tweets, and the overall campaign means, are close to the origin, i.e.\
the global average.  That just means that they, respectively -- initiating tweets, 
and means -- are relatively unexceptional, and express aggregates.  
The very low rates of inertia explained by 
the factors is an aspect which is fairly standard for such analysis of very sparse 
cross-tabulations, although it does point to the fact that we are seeing in Figure
\ref{fig1} just a projection of our data.  

Therefore, while tweets initiating campaigns 3, 5, 8 are the closest to their 
respective campaign means, this is based on the best fitting planar, two-dimensional 
dimensions.   It is based on the best factor plane, defined by factors 1 and 2.
But the entire semantic space is of dimensionality 338.  (This is explained as follows.  
The principal row set is 968 tweets.  The principal column set is 
339 tweets.  The dimensionality of the factor space is, at most and here equal to 
min($339 - 1, 968 - 1$).  Cf.\ Appendix B.)  

\begin{figure}
\centering
\includegraphics[width=14cm]{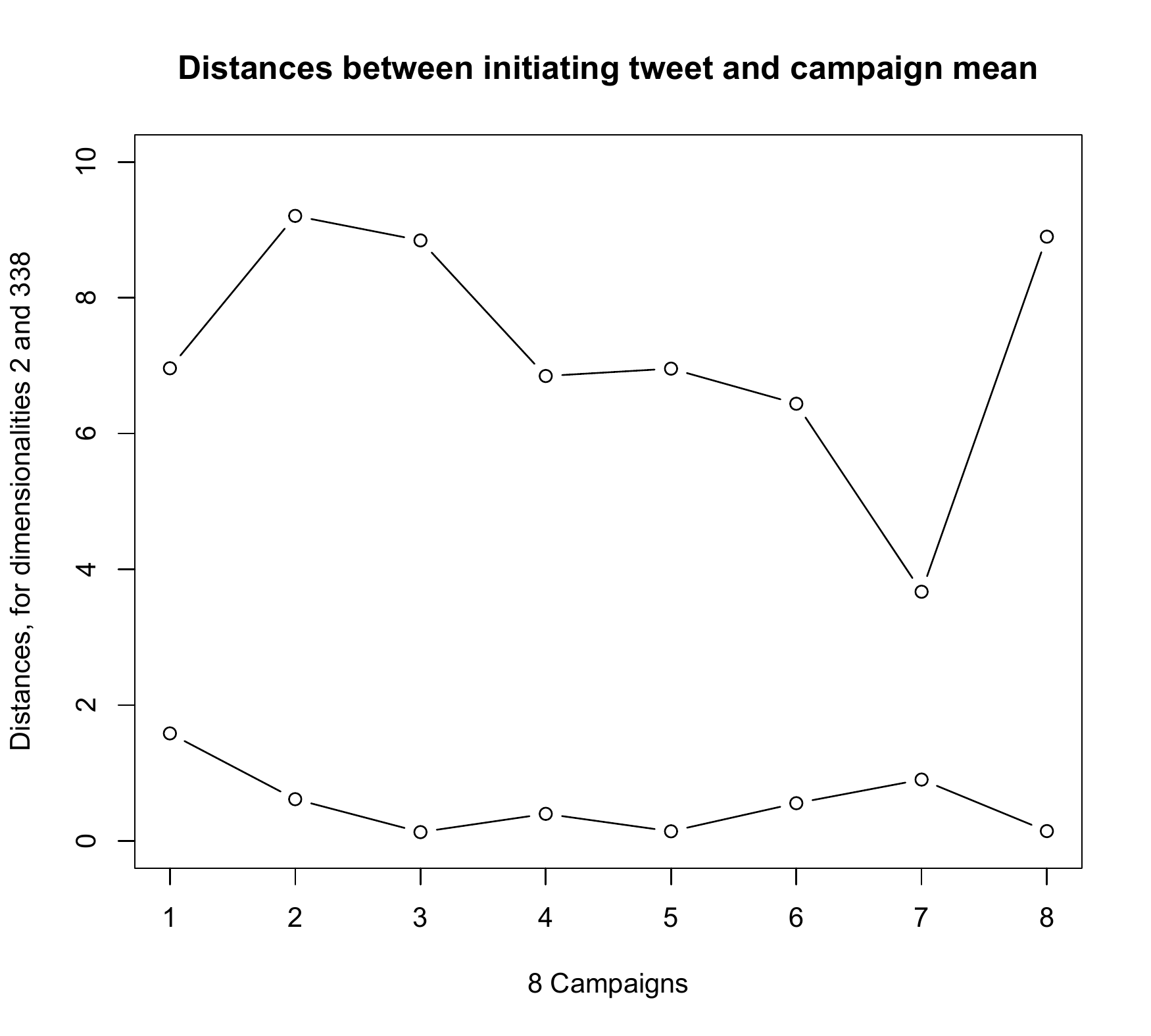}
\caption{For the 8 campaigns, shown are the Euclidean distances between the campaign 
initiating tweets and the respective centres of gravity of the campaigns, or net 
overall campaigns. The lower curve is for the principal factor plane, hence the 
Euclidean distances between ``tic1'' and ``C1'', etc., as shown in Figure \ref{fig1}.
The upper curve is for the full semantic, factor space dimensionality.}
\label{fig2}
\end{figure}

Looking at the distances 
between tweets initiating campaigns 1 to 8, relative to their respective campaign means 
(all in the factor space of dimensionality 338), see Figure \ref{fig2}, 
we find a different (and more complete) 
perspective, where campaign 7 shows the most impact by its initiating tweet, followed by 
campaigns 6, then 4, then 5, then 1.    Increasingly less impactful are 3, 8 and 2. 

\subsubsection{Statistical Significance of Impact}

We are still considering Figure \ref{fig2}.  

The campaign 7 case, with the distance between the tweet initiating campaign 7, and 
the mean campaign 7 outcome, in the full, 338-dimensional factor (semantic) space 
is equal to 3.670904. 

Compare that to all pairwise distances of non-initiating tweets.  (They are quite 
normally/Gaussian 
distributed, with a small number of large distances.)  The mean, mean $-$ stdev,
 and mean $- 2$*stdev (``stdev'' is the standard deviation) of these pairwise distances 
are: 12.64907, 8.508712, 4.368352.

We find for campaign 7, the distance between initiating tweet and mean outcome, in terms 
of the mean and stdev of all (non-initiating) tweet, full dimensionality, pairwise 
distances, to be: mean $-2.168451$*stdev

For $z = -2.16$, the campaign 7 impact is significant at the 1.5\% level 
(i.e.\  $z = -2.16$, in the two-sided case, has 98.5\% of the Gaussian greater than 
it in value).    

In the case of campaigns 1, 4, 5, 6, we find them less than 90\% of all pairwise 
distances.  

In the case of campaigns 3 and 8, we find them less than 80\% of all pairwise distances.
   
That only leaves campaign 2 as being the least good fit, relative to initiating tweet 
and outcome.   

\subsubsection{Detailed Look at Campaign 7}

Having found campaign 7 to be the best, in the full semantic dimensionality context, 
and hence with no loss whatsoever of information contained in our original data, 
from the point of view of proximity of cause and intended effect, we now look in 
somewhat more detail at this campaign. 

Campaign 7 relates to Sustainable Water use, cf.\ Appendix A. 
Including the initiating tweet, there are 112 tweets (that have not become empty of 
terms in our term filtering preprocessing) in campaign 7, and there are 176 terms that 
appear at least once in the set of tweets.  We now use Correspondence Analysis on 
just this campaign 7 data.  

\begin{figure}
\centering
\includegraphics[width=14cm]{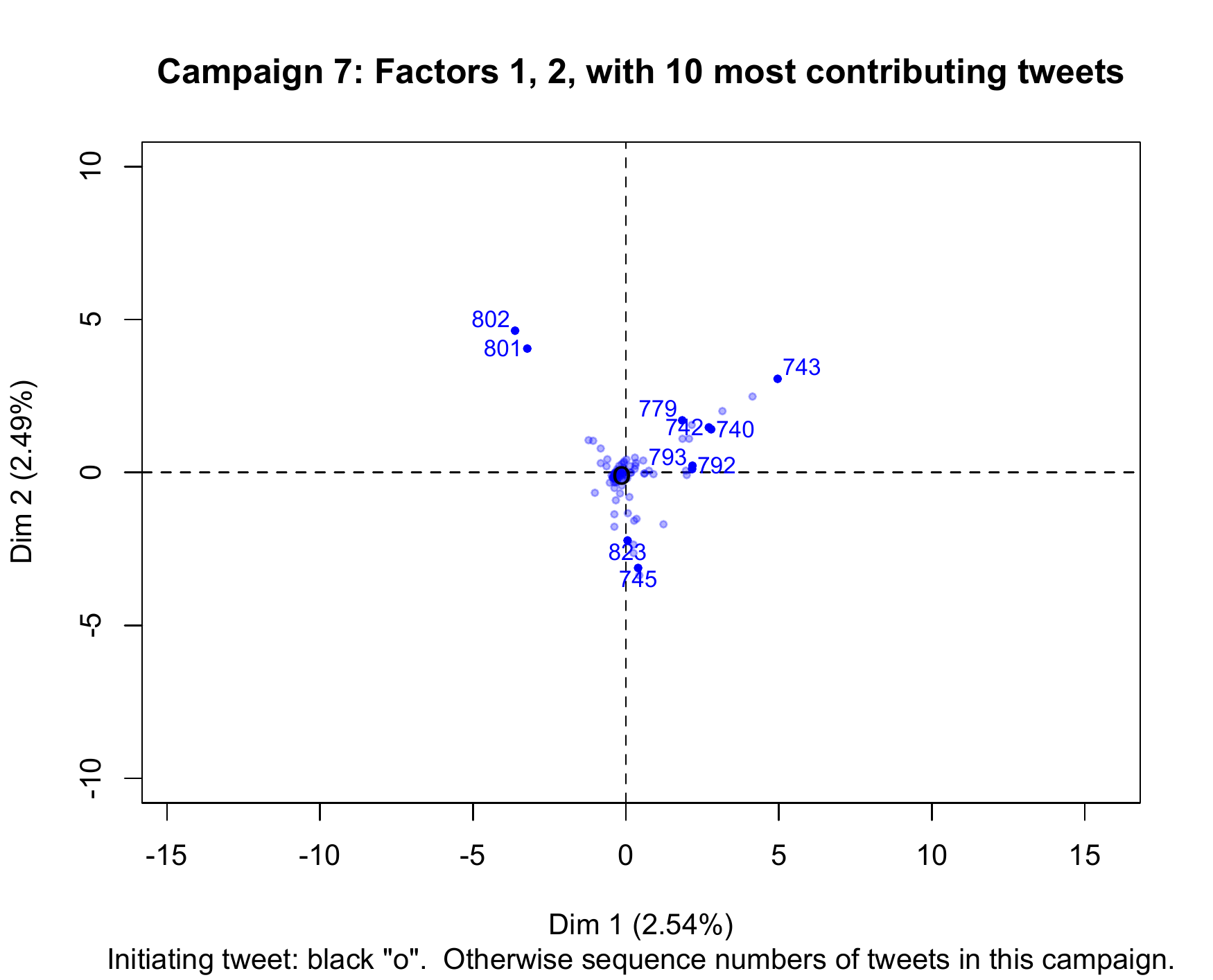}
\caption{Principal factor plane for campaign 7.  Just the tweet set for this campaign 
is used, including the initiating tweet.  Terms are used that appear at least once in 
the set of tweets.  The input data used is 112 tweets crossed by 176 terms.  The 
10 most contributing tweets are labelled here, and the initiating tweet is also 
displayed.}
\label{fig3}
\end{figure}

\begin{figure}
\centering
\includegraphics[width=14cm]{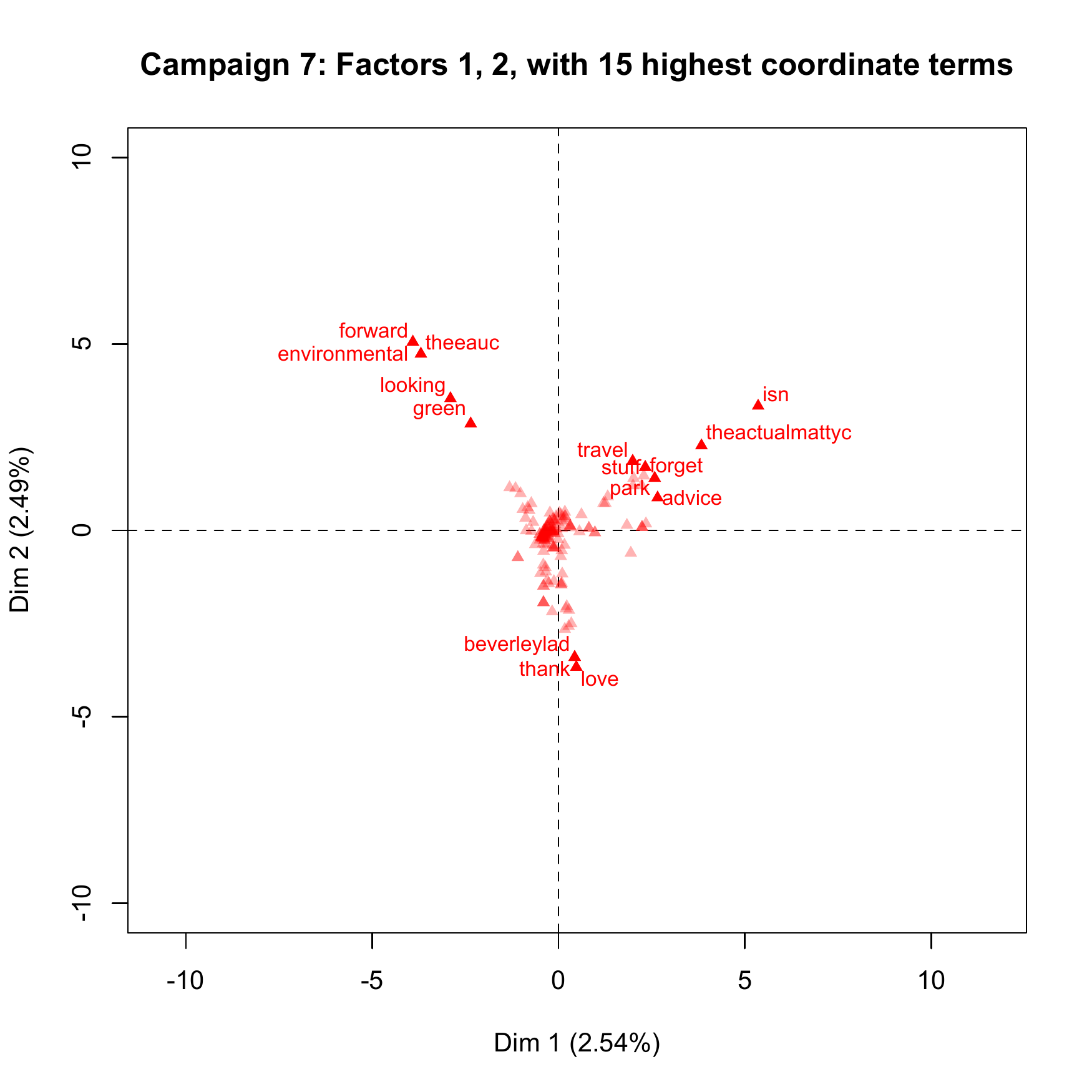}
\caption{The same data is used as in Figure \ref{fig3}.  The 15 highest coordinate
values of terms are labelled here.}
\label{fig4}
\end{figure}

We show the factors 1, 2 plane with the tweets, noting where the initiating 
tweet is located in this projection, see Figure \ref{fig3}; and then we show the 
most important terms, see Figure \ref{fig4}.  
In the latter, note the locations of tweeter names, {\tt @TheActualMattyC}, 
{\tt @TheEAUC}, {\tt @BeverleyLad}.  

The story narrated by the principal plane view of campaign 7 is very largely a 
three-way interplay of tweeter personalities, {\tt @TheActualMattyC}, {\tt @TheEAUC},
{\tt @BeverleyLad}.  Note how they are reduced in our preprocessing (cf.\ Figure 
\ref{fig4}) to, respectively, ``theactualmattyc'', ``theeauc'' and ``beverleylad''.
Respectively these are associated with: positive F1, positive F2; negative F1, 
positive F2; and relatively neutral F1, negative F2 (where F1 and F2 are factor 
1 and factor 2 coordinates).  Regarding the last of these tweeter individuals, 
the term ``love'' appears in a tweet indicating ``we'd love a cycling Leicester'',
and the word ``thanks'' appears quite a few times.  Our semantic analysis has 
provided the words shown in Figure \ref{fig4} as the most semantically loaded, 
in the factor 1, factor 2 planar projection.  

In summary, Figures \ref{fig3} and \ref{fig4} are a particular illustration of
 what campaign 7 entails.  These two figures are related to the one and the same
analysis, and are presented here as two figures in order not to have too much 
overcrowding of projections.  The information content in this planar projection 
is just over 5\% (i.e.\ 2.54\% + 2.49\%) of the total information of the campaign 
7 Twitter data.  Information is quantified by inertia explained by these factors. 
While the most important planar projection, just one twentieth of the data's 
information is quite weak.  Figures \ref{fig3} and \ref{fig4} do provide us with 
a visualization of a particular narrative underlying campaign 7.

It may be noted that in our earlier work relating to impact of a causal 
communicative action (the initiating tweet) relative to the evolution of the 
discourse (the tweets), we used the full information space, i.e.\ the full 
dimensionality of the semantic, factor space, in order to draw conclusions.

\section{Conclusions}

In the Stephen Fry Twitter case, we saw how we could visualize the critical 
tweets as a culmination of some relatively homogeneous preceding tweets, and with 
the following two tweets being semantically very different.  Hence these
following tweets manifest the shock effect. This visualization was in Figures
\ref{Fry2} and \ref{Fry2b}. 

Through segmentation of the Twitter overall narrative, we developed sub-narratives. 
These can be determined in such a way as to be statistically significant.  We discussed
the overall flow of the narrative in a way that was between the extremes (of course, 
to be checked for in the context of the given data) of being innocuous versus being
exceptional.  We took being innocuous as having a close-to-average semantic profile, in 
terms of projection in the factorial, latent semantic space.   

In the environmental citizenship experimental case study, we 
developed a new approach to assessing impact, based on the process of 
discourse.  A causal element is used, and this is compared to the overall
aggregate of a selected part (since that will be meanginful) 
of the course of the discourse.

We studied this comparatively, using 8 different ``campaigns''.
We traced out the semantic path from initiating tweet to the mean tweet of 
the associated campaign.  We noted the differences between campaigns. 
We did this using the most salient -- the most important -- two-dimensional latent 
semantic, or factor, subspace, in order to illustrate our approach, as well as in the 
full dimensionality space, using all information and avoiding any approximation.
   
We noted differences, e.g.\ campaign 3 overall was closest to its initiating tweet 
in the two-factor projection; but with all information in use, campaign 7 was the most 
effective campaign of all, in the sense of the initiating tweet being closest to the 
overall semantic mean of that campaign.    

We have also developed a statistical test of significance of impact. 

Planar projections in our semantic, factor space allow visualization of outcomes.
We looked in detail at campaign 7, pointing to what were the most influential 
tweets, and the most revealing terms associated with the underlying (latent
semantic) components.  In some cases, this indicated who (the tweeter, @) 
or what themes (hashtag, \#) were dominant, and in other cases particular words
were at issue.  

Our word sets used were carefully selected.  Nonetheless they were flexibly open 
to various grammatical forms, and to stumps of words serving as proxies for words 
containing punctuation, web addresses, or other non-standard character strings.
We avoided punctuation (including multiple explanation marks) and special characters;
we treated words that were run together as a composite word (possibly to be removed
in our preprocessing as a rare word); we avoided numeric data as being non-interpretable
for the type of narrative that is of interest to us here; and we allowed URLs or 
abbreviations to be retained in our analysis as stumps of words (again possibly 
removed due to rarity).  

In our second case study, we noted 
how top-down, command or managerially imposed approaches 
to behaviour change have been found to be often inadequate and ineffective.  Our 
motivation was to accept a Habermasian view that democratic, inclusive engagement through 
communicative processes is a better way to bring about behaviour change.  Our 
approach to quantifying impact is in this context of being process-based and 
data-driven.


\section*{Appendix A: Our 8 Campaign Initiating Tweets} 

The following are these tweets, in full.   
For campaign 4 the two initiating tweets were merged together.  DMU stands for 
De Montfort University.  

\begin{description}
\item{Campaign 1:} 
Introducing \#climatechange! Is the climate changing?What are the observed 
changes?Are humans causing it? Discuss http://t.co/cMUOmbEt \#dmuCC
\item{Campaign 2:}
Do you feel \#climatechange is a distant issue? Read and listen to the climate 
witnesses in the UK http://t.co/FLWaTqTb
\item{Campaign 3:}
Goodmorning \#DMU!! How was your weekend? Did you participate in the \#marathon? 
We are talking about electricity this week! \#dmuelectricity
\item{Campaign 4:}Goodmorning \#DMU!! How was your weekend? We are talking about gas 
and heating this week! \#dmuenergy Wishing you all a nice \#ecomonday!
\item{Campaign 4:}
Connect with us to discover what \#DMU is already doing to cut its \#gas use and 
tell us what you think we could all do to make it better!
\item{Campaign 5:}
Goodmorning \#DMU!! We talk about \#sustainable food this week. We have a question 
for you! What do you think does Sustainable Food mean?
\item{Campaign 6:}
Here I am, fueled with caffeine! This week we will be talking in particular of 
\#transport. How do you get from home to \#DMU? \#dmutransport
\item{Campaign 7:} New post! \#Sustainable \#Water | Are you familiar with the concept 
of \#WaterSecurity?  http://t.co/T9QYVlTJ \#DMU \#climate \#sustainabledmu
\item{Campaign 8:} @SustainableDMU \#MeatFreeMonday seems to have latched itself 
into my brain! Not a big meat eater but like having a dedicated veggie day!
\end{description}

As discussed in subsections \ref{preprocess1}, \ref{preprocess2}, a set of 339 terms was
ultimately selected as the set of all employable words used in the discourse.  

The terms retained for these particular initiating tweets, with frequency of occurrence,
are as follows.  For campaigns 1 through 8, we see that we have, respectively, 
summed frequencies of occurrence of terms: 4,4,7,14,10,6,7,5. 

\begin{description}
\item{Campaign 1:} climate climatechange dmucc http (all 1)
\item{Campaign 2:} climate climatechange http read (all 1)
\item{Campaign 3:} dmu electricity goodmorning participate  talking week weekend (all 1)
\item{Campaign 4:} cut dmu dmuenergy ecomonday gas goodmorning heating nice talking  tell week weekend (dmu, gas: 2; otherwise 1)
\item{Campaign 5:} dmu food goodmorning mean question sustainable talk week (food, sustainable: 2; otherwise 1)
\item{Campaign 6:} dmu dmutransport home talking transport week (all 1)
\item{Campaign 7:} climate dmu http post sustainable sustainabledmu water (all 1)
\item{Campaign 8:} day meat meatfreemonday sustainabledmu veggie (all 1)
\end{description}
Note that the campaign 4 tweet was a merged one (from original tweets 303, 304).   
In campaign 4, note that the term ``gas'' is both word and hashtag.

It is, in many cases, fairly easy to go back to the original tweets and see the 
hashtags, or the tweeters.     

We keep ``http'', although a URL originally, and it informs us 
that more information -- the web address -- is in the tweet.


\section*{Appendix B: Correspondence Analysis}

Correspondence Analysis
provides access to the semantics of information expressed by the data.
The way it does this is by viewing each observation (a tweet here) or row vector as the
average of all attributes (term here) that are related to it; and by viewing each
attribute or column vector as the average of all observations that are
related to it.

This semantic mapping analysis is as follows:

\begin{enumerate}
\item The starting point is a matrix that cross-tabulates the dependencies,
e.g.\ frequencies of joint occurrence, of an observations crossed by attributes
matrix.
\item By endowing the cross-tabulation matrix with the $\chi^2$ metric
on both observation set (rows) and attribute set (columns), we can map
observations and attributes into the same space, endowed with the Euclidean
metric.
\item Interpretation is through (i) projections of observations, attributes
onto factors; (ii) contributions by observations, attributes to the inertia 
of the factors; and (iii) correlations of observations, attributes with the 
factors.   The factors are ordered by decreasing importance.
\end{enumerate}

Correspondence Analysis
is not unlike Principal Components Analysis
in its underlying geometrical bases.
While Principal Components Analysis  is particularly
suitable for quantitative data,
Correspondence Analysis  is appropriate for the following types
of input data:
frequencies, contingency tables, probabilities, categorical data, and
mixed qualitative/categorical data.
The factors are defined by a new orthogonal coordinate system endowed with the
Euclidean distance.  The factors are determined from the eigenvectors of a 
positive semi-definite matrix (hence with non-negative eigenvalues).  This 
matrix which is diagonalized (i.e.\ subjected to singular value decomposition) 
encapsulates the requirement for the new coordinates to successively 
best fit the given data.  


The ``standardizing'' inherent in Correspondence Analysis
 (a consequence of the $\chi^2$
distance) treats rows and columns in a symmetric manner.
One byproduct is that the row and column projections in the new space
may both be plotted on the same output graphic presentations (the principal 
factor plane given by the factor 1, factor 2 coordinates; and other pairs
of factors).

\subsection*{From Frequencies of Occurrence to Clouds of Profiles, each 
Profile with an Associated Mass}

From the initial frequencies data matrix, a set of probability data,
$f_{ij}$, is defined by dividing each value by the grand total of all
elements in
the matrix.  In Correspondence Analysis,
each row (or column) point is considered to have an
associated weight.  The weight of the $i$th row point is given
by $f_i = \sum_j f_{ij}$ and the weight of the $j$th column point
is given by $f_j = \sum_i f_{ij}$. We consider the row points to have
coordinates ${f_{ij} / f_i}$, thus allowing points of the same
{\em profile} to be identical (i.e.\ superimposed).  The $i$th 
point -- because it is what we analyze -- ${f_{ij} / f_i}$ is viewed
as the conditional (empirical) probability of column $j$ given row 
$i$; and symmetrically for ${f_{ij} / f_j}$, the conditional (empirical) 
probability of row $i$ given column $j$.  

The following weighted
Euclidean distance, the $\chi^2$ distance, is then used between row
points:

\begin{equation}
d^2(i,k) = \sum_j {1 \over f_j} \left( {f_{ij} \over f_i} - 
{f_{kj} \over f_k} \right)^2
\label{eqn1} 
\end{equation}
and an analogous distance is used between column points.

The mean row point is given by the weighted average of all row
points:
\begin{equation}
\sum_i f_i {f_{ij} \over f_i} = f_j
\label{eqn2}
\end{equation}
for $j = 1, 2, \dots, m$.  Similarly the mean column profile has
$i$th coordinate $f_i$.






\subsection*{Input: Cloud of Points Endowed with the Chi Squared Metric}
The cloud of points consists of the couples:
(multidimensional) profile coordinate and (scalar) mass.
The cloud of row points, $N_I$, is the set of all $ 1 \leq i \leq n$ couples 
$ ( \{ f_{ij}/f_i | j = 1, 2, \dots p \}, f_i ) $.   
The cloud of column points, $N_J$, is the set of all $1 \leq j \leq p$ couples
$ ( \{ f_{ij}/f_j | i = 1, 2, \dots n \}, f_j ) $.   
The vectors are real-valued, so $\{ f_{ij}/f_i | j = 1, 2, \dots p \} \in \R^p$
and $\{ f_{ij}/f_j | i = 1, 2, \dots n \} \in \R^n$.

The overall inertia about the origin 
of cloud $N_I$ is: 
$$M^2(N_I) = 
\sum_i f_i \sum_j {1 \over f_j} \left( {f_{ij} \over f_i} - f_j \right)^2
= \sum_{i, j} {f_i \over f_j} \left( f_{ij} - f_i f_j \over f_i \right)^2
= \sum_{i, j} { ( f_{ij} - f_i f_j )^2 \over f_i f_j } $$
Note how this uses the $\chi^2$ distance, defined above, and how the inertia
is formally similar to the $\chi^2$ statistic of independence of observed $f_{ij}$ 
values relative to the model, $f_i f_j$, 
that is the product of the marginal probabilities. 

Similarly we have the overall inertia about the origin of cloud $N_J$: 
$$M^2(N_J) = 
\sum_j f_j \sum_i {1 \over f_i} \left( {f_{ij} \over f_j} - f_i \right)^2
= \sum_{i, j} {f_j \over f_i} \left( f_{ij} - f_i f_j \over f_j \right)^2
= \sum_{i, j} { ( f_{ij} - f_i f_j )^2 \over f_i f_j } $$
We have that the inertia of the row cloud, $N_I$, is identical to the inertia
of the column cloud, $N_J$.  

Decomposing the moment of inertia of the cloud $N_I$, or
of $N_J$ since both analyses are inherently and integrally related, furnishes the
principal axes of inertia, defined from a singular value decomposition.

\subsection*{Output: Cloud of Points Endowed with the Euclidean Metric in Factor Space}

The $\chi^2$ distance between rows $i$ and $k$, $d^2(i,k)$, has been defined in equation 
\ref{eqn1}.   
In the factor space this pairwise distance is identical, i.e.\ it is invariant.  
The coordinate
system and the metric change.  For factors indexed by $s$ and for
total dimensionality $S$, we have $ S \leq \mbox{ min } \{ n - 1, p - 1 \}$ (there 
are $n$ rows and $p$ columns);
the subtraction of 1 is since the factor space is centred and
hence there is a linear dependency which
reduces the inherent dimensionality by 1), we have the projection of
row $i$ on the $s$th factor, $F_s$, given by
$F_s(i)$:

\begin{equation}
d(i,k) = \sum_{s=1}^S \left( F_s(i) - F_s(k) \right)^2
\label{eqn3}
\end{equation}

In Correspondence Analysis the factors are ordered by decreasing
moments of inertia.  The factors are closely related, mathematically,
in the decomposition of the overall cloud,
$N_I$ and $N_J$, inertias, $M^2(N_I)$, $M^2(N_J)$.  The eigenvalues associated with the
factors, identically in the space of rows or observations indexed by set $i =
1, 2, \dots , n$,
and in the space of attributes indexed by set $j = 1, 2, \dots , p$, are given by the
eigenvalues associated with the decomposition of the inertia.  The
decomposition of the inertia is a
principal axis decomposition, which is arrived at through a singular
value decomposition.

In addition to {\em projections} on the factorial axes, for point $i$,
$F_s(i)$, and for point $j$, $G_s(j)$, we also have the following 
that are important for interpretation of results.

We have 
{\em contributions}: $f_i F^2_s(i)$ is the absolute contribution of point $i$ to the
moment of inertia $\lambda_s$, associated with factor $s$.  Contributions 
are what determine the factors or axes. 

We have also {\em correlations}.  The correlation of a point with a factor is 
the cosine squared of that point/vector with the factor/axis.  
$\cos^2 a = F^2_s(i) / \sum_{s=1}^S F^2_s(i) $ 
is the relative contribution of
the factor $s$ to point $i$.  The correlation is said to 
be the extent to which point $i$ illustrates (or exemplifies) the factor.  

Relations for column points, $j$, and factors $G_s(j)$, hold symmetrically.  

\subsection*{Analysis of the Dual Spaces, Transition Formulae, and Supplementary
Elements}

The factors in the two spaces, of rows/observations and of columns/attributes, are
inherently related as follows:

$$ F_s(i) = \lambda^{-\frac{1}{2}}_s \sum_{j=1}^p
{f_{ij} \over f_i} G_s(j) \mbox{  for  } s = 1, 2, \dots , S; i = 1, 2, \dots , n$$
\begin{equation}
G_s(j) = \lambda^{-\frac{1}{2}}_s \sum_{i=1}^n
{f_{ij} \over f_j} F_s(i) \mbox{  for  } s = 1, 2, \dots , S; j = 1, 2, \dots , p
\label{eqn4}
\end{equation}

These are termed the {\em transition formulas}.  The coordinate of
element $i$, $1 \leq i \leq n$, is the barycentre (centre of gravity)  
of the coordinates of the elements
$j$, $1 \leq j \leq p$, with associated masses of value given by the coordinates of
$f_{ij}/f_i$ of the profile of $i$.  This is all to within the
$\lambda^{-\frac{1}{2}}_s$ constant.

We can consider normalized factors, 
$ \phi_s(i) = \lambda^{-\frac{1}{2}}_s F_s(i)$, and similarly 
$ \psi_s(j) = \lambda^{-\frac{1}{2}}_s G_s(j)$. 
 
Therefore
$$ \phi_s(i) = \sum_{j=1}^p
{f_{ij} \over f_i} \psi_s(j) \mbox{  for  } s = 1, 2, \dots , S; i = 1, 2, \dots , n$$
\begin{equation}
\psi_s(j) = \sum_{i=1}^n
{f_{ij} \over f_j} \phi_s(i) \mbox{  for  } s = 1, 2, \dots , S; j = 1, 2, \dots , p
\label{eqn5}
\end{equation}

This implies that we can pass easily from one space to the other.  
us to simultaneously view and interpret observations and attributes.

Qualitatively different elements (i.e.\ row or column profiles), or
ancillary characterization or descriptive elements 
may be placed as {\em supplementary elements}.  This
means that they are given zero mass in the analysis, and their projections
are determined using the transition formulas.  This amounts to carrying
out a Correspondence Analysis first, without these elements, and then
projecting them into the factor space following the determination of all
properties of this space.

The transition formulas allow {\em supplementary rows} or columns to be
\index{supplementary element}
projected into either space.  If $\xi_{j}$ is the $j$th element of
a supplementary row, with mass $\xi$, then a factor loading, for factor $s$, is simply
obtained subsequent to the Correspondence Analysis:
$$ \psi_i = {1 \over \sqrt{\lambda}} \sum_j {\xi_{j} \over \xi} \phi_j . $$

A similar formula holds for supplementary columns.  Such
supplementary elements
are therefore ``passive'' and are incorporated into the
Correspondence Analysis results subsequent
to the eigen-analysis being carried out.

\subsection*{In Summary}

Correspondence Analysis is thus the inertial decomposition of the dual clouds of 
weighted points.   It is a latent semantic decomposition, where the 
role of the term frequency and
inverse document frequency (TF-IDF) weighting scheme is instead the use of (i) profiles 
and masses, (ii) with the $\chi^2$ distance.  See \cite{saporta} for a discussion of
both methods, Correspondence Analysis and Latent Semantic Indexing. 

Further background description can be found in \cite{benz,gopalan,leroux,ref08888}.


\begin{thebibliography}{99}

\bibitem{barker}
Barker M, Barker DI, Bormann NF, Neher KE: {\em Social Media 
Marketing. A Strategic Approach}, Andover UK: Cengage Learning; 2012.


\bibitem{becue}
B\'ecue-Bertaut M, Kostov B, Morin A, Naro G: 
Rhetorical strategy in forensic speeches: Multidimensional statistics-based 
methodology, {\em Journal of Classification}, 2014, 31:85--106.

\bibitem{benz}
Benz\'ecri, J-P: {\em L'Analyse des Donn\'ees, Tome I Taxinomie,
Tome II Correspondances}, 2nd ed.\ Paris: Dunod; 1979.

\bibitem{gopalan}
Benz\'ecri J-P: {\em Correspondence Analysis Handbook}, Basel: Dekker; 1994.  


\bibitem{greenacre}
Blasius J, Greenacre M (Eds): {\em Visualization and Verbalization of Data},
Boca Raton, FL: Chapman \& Hall/CRC Press; 2014.



\bibitem{chafe1}
Chafe WL: The flow of thought and the flow of language, in Giv\'on T (Ed),
{\em Syntax and
Semantics: Discourse and Syntax (Vol. 12)}, New York: Academic
Press, 1979, pp. 159--181.

\bibitem{chafe2}
Chafe W: {\em Discourse, Consciousness, and Time. The Flow and Displace-
ment of Conscious Experience in Speaking and Writing}, Chicago: University of
Chicago Press; 1994.

\bibitem{chew}
Chew C, Eysenbach G: Pandemics in the age of Twitter: content analysis of tweets
during the 2009 H1N1 outbreak, {\em PLoS ONE}, 2010, 5(11):e14118, 13 pp..  


\bibitem{finnis}
Finnis F, Chan S, Clements R: Let's get real. How to Evaluate Online
Success? Report from the Culture24 Action Research Project. Brighton, 40 pp., 2011
(http://www.keepandshare.com/doc/3148918/
culture24-howtoevaluateonlinesuccess-2-pdf-september-19-2011-11-15-am-2-5-meg?da=y
retrieved 13 April 2014). 


\bibitem{howto}
Howto.gov, Social Media Metrics for Federal Agencies, U.S. General Services
Administration, 2013
(http://www.howto.gov/social-media/using-social-media-in-government/metrics-for-federal-agencies retrieved 13 April 2014). 

\bibitem{legendre}
Legendre P, Legendre L: {\em Numerical Ecology}, (3rd ed), Amsterdam: Elsevier; 
2012.

\bibitem{leroux}
Le Roux B, Rouanet H: 
{\em Geometric Data Analysis: From Correspondence Analysis to Structured
Data Analysis}, Dordrecht: Kluwer Academic; 2004.


\bibitem{mckee}
McKee R: {\em Story: Substance, Structure, Style, and the Principles of Screenwriting}.
York: Methuen; 1999.  

\bibitem{ref08888}
Murtagh F: {\em Correspondence Analysis and Data Coding with R and Java}.
Boca Raton,FL: Chapman \& Hall/CRC; 2005.

\bibitem{murtaghganzmckie}
Murtagh F, Ganz A, McKie S: The structure of narrative: The case
of film scripts, {\em Pattern Recognition}, 2009, 42: 302--312.

\bibitem{murtaghganzreddington}
Murtagh F, Ganz A, Reddington J: New methods of analysis of
narrative and semantics in support of interactivity, {\em Entertainment Computing}, 
2011, 2: 115--121.

\bibitem{pearce}
Pearce W, Holmberg K, Hellsten I, Nerlich B: Climate change on Twitter: 
topics, communities and conversations about the 2013 IPCC Working Group 1 report,
{\em PLoS ONE}, 9(4), 11 pp., e94785, 2014. 

\bibitem{pianosi2013}
Pianosi M, Bull R, Rieser M: Impact, influence and reach: Lessons in 
measuring the impact of social media, pp. 36, preprint, 2013.

\bibitem{quinn}
Quinn B: Stephen Fry fans beg actor not to give up on Twitter, 
{\em The Observer}, 31 October 2009.
http://www.theguardian.com/technology/2009/oct/31/stephen-fry-leave-twitter-fans

\bibitem{saporta}
S\'egu\'ela J  and Saporta G:   
A comparison between latent semantic analysis and correspondence
analysis, presentation, {\em CARME, Correspondence Analysis and Related Methods 
Conference}, Rennes, France, 2011.
http//carme2011.agrocampus-ouest.fr/slides/Seguela\_Saporta.pdf.



\end{thebibliography}
\end{document}